\pdfoutput=1
\documentclass[12pt,a4paper]{article}

\usepackage{ifthen} 
\newboolean{pdflatex}
\setboolean{pdflatex}{true} 

\newboolean{articletitles}
\setboolean{articletitles}{true} 

\newboolean{uprightparticles}
\setboolean{uprightparticles}{true} 

\newboolean{completeplots}
\setboolean{completeplots}{false}

\usepackage[dvipsnames]{xcolor}

\def\paperauthors{LHCb collaboration} 
\def\paperasciititle{} 
\def\papertitle{Study of charmonium and charmonium-like contributions in~\mbox{$\decay{\Bp}{\jpsi\Peta\Kp}$}~decays} 
\def\paperkeywords{{High Energy Physics}, {LHCb}} 
\def\papercopyright{\the\year\ CERN for the benefit of the LHCb collaboration} 
\def\paperlicence{CC BY 4.0 licence}
\def\paperlicenceurl{https://creativecommons.org/licenses/by/4.0/}

\usepackage[top=1in, bottom=1.25in, left=1in, right=1in]{geometry}

\usepackage{microtype}
\usepackage[mathlines]{lineno}  
\usepackage{xspace} 
\usepackage{caption} 

\usepackage{graphicx}  
\usepackage{color}
\usepackage{colortbl}
\graphicspath{{./figs/}} 

\usepackage{amsmath} 
\usepackage{amssymb}
\usepackage{amsfonts}
\usepackage{upgreek} 

\newcommand*\patchAmsMathEnvironmentForLineno[1]{%
\expandafter\let\csname old#1\expandafter\endcsname\csname #1\endcsname
\expandafter\let\csname oldend#1\expandafter\endcsname\csname
end#1\endcsname
 \renewenvironment{#1}%
   {\linenomath\csname old#1\endcsname}%
   {\csname oldend#1\endcsname\endlinenomath}%
}
\newcommand*\patchBothAmsMathEnvironmentsForLineno[1]{%
  \patchAmsMathEnvironmentForLineno{#1}%
  \patchAmsMathEnvironmentForLineno{#1*}%
}
\AtBeginDocument{%
\patchBothAmsMathEnvironmentsForLineno{equation}%
\patchBothAmsMathEnvironmentsForLineno{align}%
\patchBothAmsMathEnvironmentsForLineno{flalign}%
\patchBothAmsMathEnvironmentsForLineno{alignat}%
\patchBothAmsMathEnvironmentsForLineno{gather}%
\patchBothAmsMathEnvironmentsForLineno{multline}%
\patchBothAmsMathEnvironmentsForLineno{eqnarray}%
}

\usepackage{hyperxmp}

\usepackage[pdftex,
            pdfauthor={\paperauthors},
            pdftitle={\paperasciititle},
            pdfkeywords={\paperkeywords},
            pdfcopyright={Copyright (C) \papercopyright},
            pdflicenseurl={\paperlicenceurl}]{hyperref}

\usepackage[colorinlistoftodos,textsize=scriptsize]{todonotes}

\usepackage[bottom,flushmargin,hang,multiple]{footmisc}

\usepackage[all]{hypcap} 

\usepackage{xspace} 
\usepackage{upgreek}

\def\lhcb   {\mbox{LHCb}\xspace}

\def\babar  {\mbox{BaBar}\xspace}
\def\belle  {\mbox{Belle}\xspace}

\def\MagUp {\mbox{\em Mag\kern -0.05em Up}\xspace}

\ifthenelse{\boolean{uprightparticles}}%
{
 
 \def\Pgamma      {\ensuremath{\upgamma}\xspace}

 \def\Peta        {\ensuremath{\upeta}\xspace}

 \def\Pmu         {\ensuremath{\upmu}\xspace}

 \def\Ppi         {\ensuremath{\uppi}\xspace}

 \def\Pchi        {\ensuremath{\upchi}\xspace}                 
 \def\Ppsi        {\ensuremath{\uppsi}\xspace}

 \def\PDelta      {\ensuremath{\Delta}\xspace}                 
 \def\PXi         {\ensuremath{\Xi}\xspace}                 
 \def\PLambda     {\ensuremath{\Lambda}\xspace}                 
 \def\PSigma      {\ensuremath{\Sigma}\xspace}                 
 \def\POmega      {\ensuremath{\Omega}\xspace}                 
 \def\PUpsilon    {\ensuremath{\Upsilon}\xspace}

 \def\PB      {\ensuremath{\mathrm{B}}\xspace}                 
 \def\PC      {\ensuremath{\mathrm{C}}\xspace}                 
 \def\PD      {\ensuremath{\mathrm{D}}\xspace}

 \def\PJ      {\ensuremath{\mathrm{J}}\xspace}                 
 \def\PK      {\ensuremath{\mathrm{K}}\xspace}

 \def\PP      {\ensuremath{\mathrm{P}}\xspace}                 
                  
 \def\PR      {\ensuremath{\mathrm{R}}\xspace}                 
 \def\PS      {\ensuremath{\mathrm{S}}\xspace}

 \def\PX      {\ensuremath{\mathrm{X}}\xspace}                 
                  
 \def\PZ      {\ensuremath{\mathrm{Z}}\xspace}                 
                  
 \def\Pb      {\ensuremath{\mathrm{b}}\xspace}                 
 \def\Pc      {\ensuremath{\mathrm{c}}\xspace}

 \def\Pg      {\ensuremath{\mathrm{g}}\xspace}                 
                  
 \def\Pi      {\ensuremath{\mathrm{i}}\xspace}

 \def\Pp      {\ensuremath{\mathrm{p}}\xspace}

 \def\Ps      {\ensuremath{\mathrm{s}}\xspace}

 \def\thebaroffset{0.0em}
}
{
 
 \def\Pgamma      {\ensuremath{\gamma}\xspace}

 \def\Peta        {\ensuremath{\eta}\xspace}

 \def\Pmu         {\ensuremath{\mu}\xspace}

 \def\Ppi         {\ensuremath{\pi}\xspace}

 \def\Pchi        {\ensuremath{\chi}\xspace}                 
 \def\Ppsi        {\ensuremath{\psi}\xspace}                 
                  
 \mathchardef\PDelta="7101
 \mathchardef\PXi="7104
 \mathchardef\PLambda="7103
 \mathchardef\PSigma="7106
 \mathchardef\POmega="710A
 \mathchardef\PUpsilon="7107
                  
 \def\PB      {\ensuremath{B}\xspace}                 
 \def\PC      {\ensuremath{C}\xspace}                 
 \def\PD      {\ensuremath{D}\xspace}

 \def\PJ      {\ensuremath{J}\xspace}                 
 \def\PK      {\ensuremath{K}\xspace}

 \def\PP      {\ensuremath{P}\xspace}                 
                  
 \def\PR      {\ensuremath{R}\xspace}                 
 \def\PS      {\ensuremath{S}\xspace}

 \def\PX      {\ensuremath{X}\xspace}                 
                  
 \def\PZ      {\ensuremath{Z}\xspace}                 
                  
 \def\Pb      {\ensuremath{b}\xspace}                 
 \def\Pc      {\ensuremath{c}\xspace}

 \def\Pg      {\ensuremath{g}\xspace}                 
                  
 \def\Pi      {\ensuremath{i}\xspace}

 \def\Pp      {\ensuremath{p}\xspace}

 \def\Ps      {\ensuremath{s}\xspace}

 \def\thebaroffset{0.18em}
}
\newcommand{\offsetoverline}[2][\thebaroffset]{\kern #1\overline{\kern -#1 #2}}%

\makeatletter
\ifcase \@ptsize \relax
  \newcommand{\miniscule}{\@setfontsize\miniscule{4}{5}}
\or
  \newcommand{\miniscule}{\@setfontsize\miniscule{5}{6}}
\or
  \newcommand{\miniscule}{\@setfontsize\miniscule{5}{6}}
\fi
\makeatother

\DeclareRobustCommand{\optbar}[1]{\shortstack{{\miniscule (\rule[.5ex]{1.25em}{.18mm})}
  \\ [-.7ex] $#1$}}

\def\mumu       {{\ensuremath{\Pmu^+\Pmu^-}}\xspace}

\def\g      {{\ensuremath{\Pgamma}}\xspace}

\def\Z      {{\ensuremath{\PZ}}\xspace}

\def\squark    {{\ensuremath{\Ps}}\xspace}

\def\cquark    {{\ensuremath{\Pc}}\xspace}
\def\cquarkbar {{\ensuremath{\overline \cquark}}\xspace}

\def\bquark    {{\ensuremath{\Pb}}\xspace}

\def\pion   {{\ensuremath{\Ppi}}\xspace}
\def\piz    {{\ensuremath{\pion^0}}\xspace}
\def\pip    {{\ensuremath{\pion^+}}\xspace}
\def\pim    {{\ensuremath{\pion^-}}\xspace}

\def\kaon    {{\ensuremath{\PK}}\xspace}

\def\KorKbar {\kern \thebaroffset\optbar{\kern -\thebaroffset \PK}{}\xspace}

\def\Kp      {{\ensuremath{\kaon^+}}\xspace}
\def\Km      {{\ensuremath{\kaon^-}}\xspace}
\def\Kpm     {{\ensuremath{\kaon^\pm}}\xspace}

\def\Kstarp  {{\ensuremath{\kaon^{*+}}}\xspace}

\def\Dbar    {{\ensuremath{\offsetoverline{\PD}}}\xspace}
\def\D       {{\ensuremath{\PD}}\xspace}

\def\DorDbar {\kern \thebaroffset\optbar{\kern -\thebaroffset \PD}\xspace}
\def\Dz      {{\ensuremath{\D^0}}\xspace}
\def\Dzb     {{\ensuremath{\Dbar{}^0}}\xspace}
\def\Dp      {{\ensuremath{\D^+}}\xspace}
\def\Dm      {{\ensuremath{\D^-}}\xspace}

\def\DpDm    {\ensuremath{\Dp {\kern -0.16em \Dm}}\xspace}

\def\Dstarz  {{\ensuremath{\D^{*0}}}\xspace}

\def\Dstarp  {{\ensuremath{\D^{*+}}}\xspace}

\def\B       {{\ensuremath{\PB}}\xspace}

\def\BorBbar {\kern \thebaroffset\optbar{\kern -\thebaroffset \PB}\xspace}

\def\Bd      {{\ensuremath{\B^0}}\xspace}

\def\BdorBdbar {\kern \thebaroffset\optbar{\kern -\thebaroffset \Bd}\xspace}
\def\Bu      {{\ensuremath{\B^+}}\xspace}

\def\Bp      {{\ensuremath{\Bu}}\xspace}

\def\Bpm     {{\ensuremath{\B^\pm}}\xspace}

\def\Bs      {{\ensuremath{\B^0_\squark}}\xspace}

\def\BsorBsbar {\kern \thebaroffset\optbar{\kern -\thebaroffset \Bs}\xspace}

\def\jpsi     {{\ensuremath{{\PJ\mskip -3mu/\mskip -2mu\Ppsi}}}\xspace}
\def\psitwos  {{\ensuremath{\Ppsi{(2\PS)}}}\xspace}

\def\chicone  {{\ensuremath{\Pchi_{\cquark 1}}}\xspace}

\def\Y#1S{\ensuremath{\PUpsilon{(#1S)}}\xspace}

\def\proton      {{\ensuremath{\Pp}}\xspace}

\def\LorLbar     {\kern \thebaroffset\optbar{\kern -\thebaroffset \PLambda}\xspace}

\def\BF         {{\ensuremath{\mathcal{B}}}\xspace}
\def\BR         {\BF}

\newcommand{\decay}[2]{\ensuremath{#1\!\to #2}\xspace} 

\def\to                 {\ensuremath{\rightarrow}\xspace}

\def\AT#1     {\ensuremath{A_{\mathrm{T}}^{#1}}\xspace}           

\def\C#1      {\ensuremath{\mathcal{C}_{#1}}\xspace}                       
\def\Cp#1     {\ensuremath{\mathcal{C}_{#1}^{'}}\xspace}                    
\def\Ceff#1   {\ensuremath{\mathcal{C}_{#1}^{\mathrm{(eff)}}}\xspace}        
\def\Cpeff#1  {\ensuremath{\mathcal{C}_{#1}^{'\mathrm{(eff)}}}\xspace}       
\def\Ope#1    {\ensuremath{\mathcal{O}_{#1}}\xspace}                       
\def\Opep#1   {\ensuremath{\mathcal{O}_{#1}^{'}}\xspace}                    

\newcommand{\nospaceunit}[1]{\ensuremath{\text{#1}}}       
\newcommand{\aunit}[1]{\ensuremath{\text{\,#1}}}       

\newcommand{\tev}{\aunit{Te\kern -0.1em V}\xspace}
\newcommand{\gev}{\aunit{Ge\kern -0.1em V}\xspace}
\newcommand{\mev}{\aunit{Me\kern -0.1em V}\xspace}
\newcommand{\kev}{\aunit{ke\kern -0.1em V}\xspace}
\newcommand{\ev}{\aunit{e\kern -0.1em V}\xspace}
 
\newcommand{\mevc}{\ensuremath{\aunit{Me\kern -0.1em V\!/}c}\xspace}
\newcommand{\gevc}{\ensuremath{\aunit{Ge\kern -0.1em V\!/}c}\xspace}
\newcommand{\mevcc}{\ensuremath{\aunit{Me\kern -0.1em V\!/}c^2}\xspace}
\newcommand{\gevcc}{\ensuremath{\aunit{Ge\kern -0.1em V\!/}c^2}\xspace}

\def\mum  {\ensuremath{\,\upmu\nospaceunit{m}}\xspace}

\def\fb   {\ensuremath{\aunit{fb}}\xspace}
\def\invfb   {\ensuremath{\fb^{-1}}\xspace}

\newcommand{\chisq}{\ensuremath{\chi^2}\xspace}

\newcommand{\chisqip}{\ensuremath{\chi^2_{\text{IP}}}\xspace}

\def\gsim{{~\raise.15em\hbox{$>$}\kern-.85em
          \lower.35em\hbox{$\sim$}~}\xspace}
\def\lsim{{~\raise.15em\hbox{$<$}\kern-.85em
          \lower.35em\hbox{$\sim$}~}\xspace}

\def\pt         {\ensuremath{p_{\mathrm{T}}}\xspace}

\def\ptot       {\ensuremath{p}\xspace}

\def\evtgen     {\mbox{\textsc{EvtGen}}\xspace}

\def\geant      {\mbox{\textsc{Geant4}}\xspace}

\def\photos     {\mbox{\textsc{Photos}}\xspace}

\def\pythia     {\mbox{\textsc{Pythia}}\xspace}

\def\tell1  {TELL1\xspace}
\def\ukl1   {UKL1\xspace}

\newcommand{\eg}{\mbox{\itshape e.g.}\xspace}

\usepackage{cite} 
\usepackage{mciteplus}

\usepackage{graphpap}
\usepackage{comment}

\usepackage{multirow} 
\usepackage{rotating} 
\usepackage{afterpage}
\usepackage{dashrule}
\usepackage{tikz}
\usetikzlibrary{patterns}
\usepackage{wasysym}
\usepackage{cancel}
\usepackage[normalem]{ulem}

\newcommand{\kevc}{\ensuremath{\aunit{ke\kern -0.1em V\!/}c}\xspace}
\newcommand{\kevcc}{\ensuremath{\aunit{ke\kern -0.1em V\!/}c^2}\xspace}

\DeclareMathOperator*{\bigplus}{\scalerel*{+}{\sum}}
\usepackage{scalerel}

\def\XXint#1#2#3{{\setbox0=\hbox{$#1{#2#3}{\int}$}
     \vcenter{\hbox{$#2#3$}}\kern-.5\wd0}}

\makeatletter
\g@addto@macro\bfseries{\boldmath}
\makeatother

\begin{document}

\renewcommand{\thefootnote}{\fnsymbol{footnote}}
\setcounter{footnote}{1}

\begin{titlepage}
\pagenumbering{roman}

\vspace*{-1.5cm}
\centerline{\large EUROPEAN ORGANIZATION FOR NUCLEAR RESEARCH (CERN)}
\vspace*{1.5cm}
\noindent
\begin{tabular*}{\linewidth}{lc@{\extracolsep{\fill}}r@{\extracolsep{0pt}}}
\ifthenelse{\boolean{pdflatex}}
{\vspace*{-1.5cm}\mbox{\!\!\!\includegraphics[width=.14\textwidth]{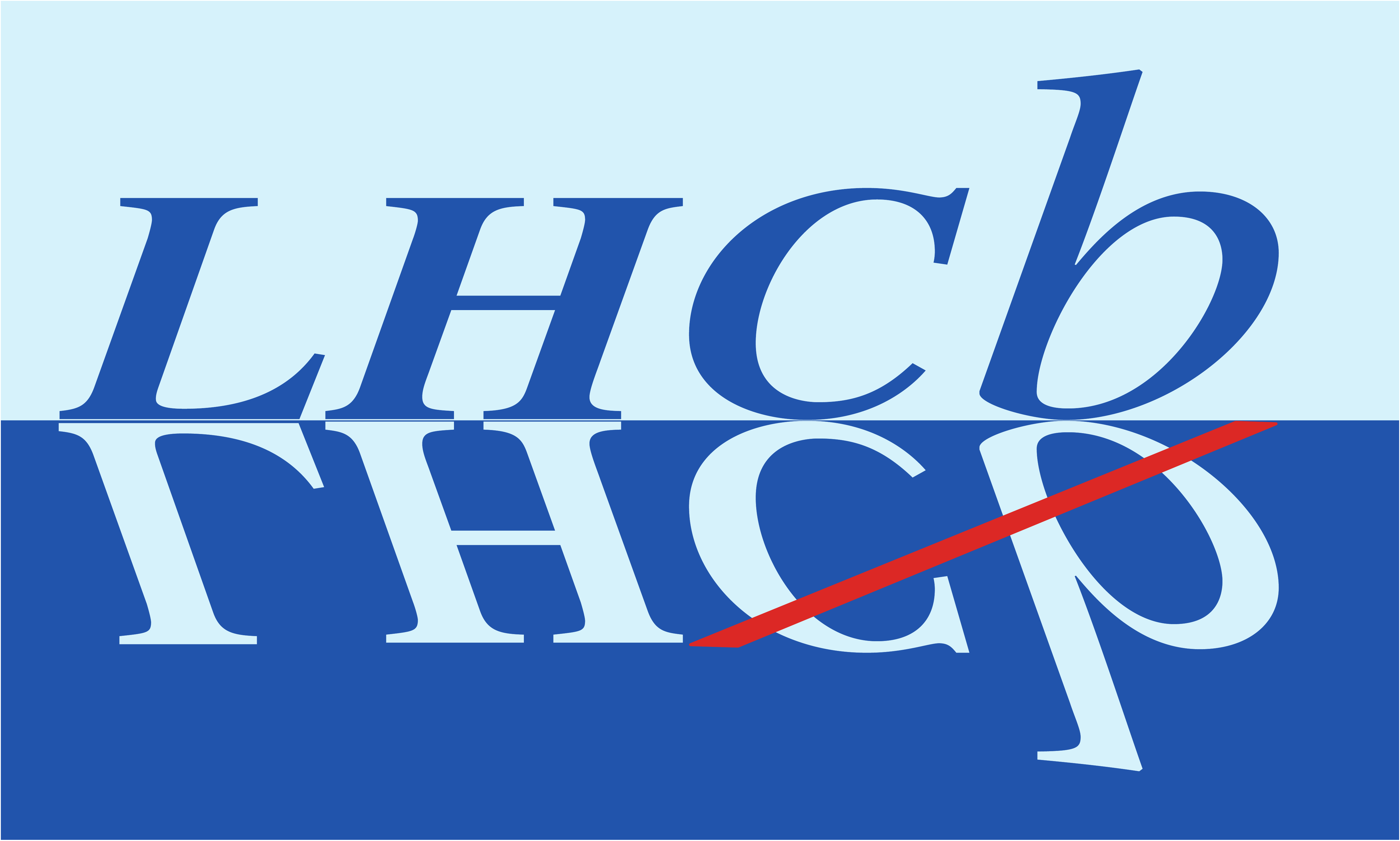}} & &}%
{\vspace*{-1.2cm}\mbox{\!\!\!\includegraphics[width=.12\textwidth]{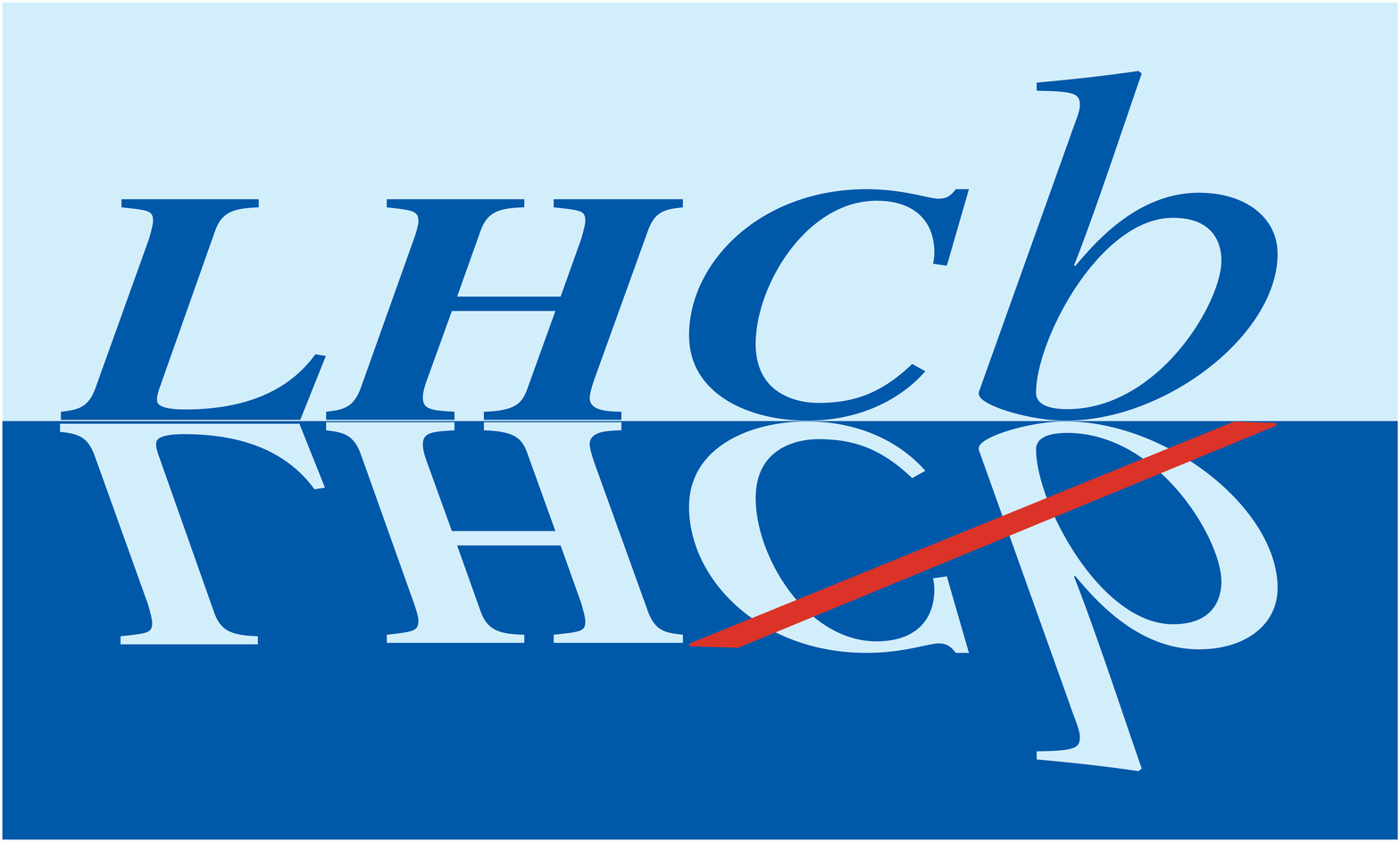}} & &}%
\\
 & & CERN-EP-2022-009 \\  
 & & LHCb-PAPER-2021-047 \\  
& & February 4, 2022 \\ 
\end{tabular*}

\vspace*{2.0cm}

{\normalfont\bfseries\boldmath\huge
\begin{center}
  \papertitle 
\end{center}
}

\vspace*{1.5cm}

\begin{center}
\paperauthors\footnote{Authors are listed at the end of this paper.}
\end{center}

\vspace{\fill}

\begin{abstract}
	\noindent
	
A study of \decay{\Bp}{\jpsi\Peta\Kp} 
decays, followed by \decay{\jpsi}{\mumu} and 
\decay{\Peta}{\g\g}, 
is performed
using a dataset collected with the LHCb detector 
in proton\nobreakdash-proton collisions 
at centre-of-mass energies
of 7, 8 and 13\tev, corresponding
to an integrated luminosity of 9\invfb. 
The~$\jpsi\Peta$~mass spectrum is investigated 
for contributions from 
charmonia and charmonium\nobreakdash-like 
states.
 Evidence is found for 
 the~\mbox{$\decay{\Bp}{\left( \decay{\Ppsi_2(3823)}{\jpsi\Peta} \right) \Kp}$}
 and \mbox{$\decay{\Bp}{\left( \decay{\Ppsi(4040)}{\jpsi\Peta} \right) \Kp}$}~decays
 with significance of 3.4 and 4.7~standard deviations, respectively. This constitutes the~first~evidence for
 the~\mbox{$\decay{\Ppsi_2(3823)}{\jpsi\Peta}$}~decay.

\end{abstract}

\vspace*{2.0cm}

\begin{center}
  Published in 
  {\href{https://doi.org/10.1007/JHEP04(2022)046}{
  JHEP {\bf{2022}}, 46 (2022)}}
\end{center}

\vspace{\fill}

{\footnotesize 
\centerline{\copyright~\papercopyright. \href{\paperlicenceurl}{\paperlicence}.}}
\vspace*{2mm}

\end{titlepage}


\newpage
\setcounter{page}{2}
\mbox{~}
%
%
%
%

\renewcommand{\thefootnote}{\arabic{footnote}}
\setcounter{footnote}{0}

\cleardoublepage

\pagestyle{plain} 
\setcounter{page}{1}
\pagenumbering{arabic}
\section{Introduction}
\label{sec:Introduction}

Exclusive $\B$\nobreakdash-meson decays provide 
an~excellent opportunity 
for studies of charmonium and 
charmonium\nobreakdash-like exotic states. 
The~enigmatic $\chicone(3872)$~particle, 
also known as~$\PX(3872)$,
was the~first discovered 
charmonium\nobreakdash-like state,
observed by the~Belle collaboration
in the~$\jpsi\pip\pim$~mass spectrum 
from \mbox{$\decay{\Bp}
{\jpsi\pip\pim\Kp}$}~decays~\cite{Belle:2003nnu}.
The~properties of this state have been extensively 
studied by 
the~CDF, D0, BaBar, Belle, LHCb, CMS, 
BESIII and ATLAS~collaborations~\mbox{\cite{Acosta:2003zx,
Abazov:2004kp,
Aubert:2004ns,
Abulencia:2005zc,
Aubert:2005zh,
Aubert:2006aj,
Abulencia:2006ma,
Aubert:2007pz,
Aubert:2008gu,
Aaltonen:2009vj,
Aubert:2008ae,
Adachi:2008sua,
delAmoSanchez:2010jr,
Choi:2011fc,
Bhardwaj:2011dj,
LHCb-PAPER-2011-034,
LHCb-PAPER-2013-001,
Chatrchyan:2013cld,
Ablikim:2013dyn,
LHCb-PAPER-2014-008,
Bala:2015wep,
LHCb-PAPER-2015-015,
Aaboud:2016vzw,
LHCb-PAPER-2016-016,
LHCb-PAPER-2019-023,
Chou:2019mlv,
Bhardwaj:2019spn,
LHCb-PAPER-2020-008,
LHCb-PAPER-2020-009,
LHCb-PAPER-2020-023,
CMS:2020eiw,
LHCb-PAPER-2020-035}.}
The~$\chicone(3872)$ particle 
was followed by observations of numerous 
charmonium\nobreakdash-like
states~\cite{PDG2021,Brambilla:2019esw}
that are incompatible 
with having 
$\cquark\cquarkbar$~quark content,  
sparking a~wave of interest  in exotic hadron
spectroscopy~\mbox{\cite{Swanson:2006st,
Chen:2016qju,
Esposito:2016noz,
Ali:2017jda,
Hosaka:2016pey,
Lebed:2016hpi,
Guo:2017jvc,
Olsen:2017bmm,
Brambilla:2019esw,
Ali:2019roi}.}
\mbox{Despite} all theoretical and experimental efforts
the~nature of such states is not yet understood.
For~instance,
the~narrow width  of
the~$\chicone(3872)$~state and 
its proximity 
to 
the~$\Dstarz\Dzb$~mass threshold~\mbox{\cite{LHCb-PAPER-2020-008,
LHCb-PAPER-2020-009}}
support
the~$\chicone(3872)$~state 
to be a~loosely 
bound $\Dstarz\Dzb$~molecule~\mbox{\cite{Swanson:2004cq,
Liu:2009qhy,
Lee:2009hy,
Ortega:2010zza,
Albaladejo:2017blx,
Kalashnikova:2018vkv,
Chen:2021iaw}.}
Other hypotheses 
include, but are not limited to, 
a~tetraquark state~\mbox{\cite{Maiani:2004vq,
Maiani:2014aja,
Maiani:2017kyi},} 
a~$\cquark \cquarkbar\Pg$~hybrid meson~\cite{hybrid_meson}, 
a~vector glueball model~\cite{AlFiky:2007zz},
a~hadro\nobreakdash-charmonium~\cite{Dubynskiy:2008mq}, 
a~cusp~\cite{Swanson:2006st}  
or
a~$\chicone(2\PP)$~charmonium state~\cite{Achasov:2015wea,Achasov:2015oia}. 
For some models, 
the~existence of $\chicone(3872)$~partner states 
is predicted~\mbox{\cite{Maiani:2004vq,
tetraquark,
Guo:2013gka,
Maiani:2014aja,
Maiani:2017kyi,
Baru:2017qwx,
Baru:2017pvh,
Mutuk:2018zxs,
Liu:2019stu}}.
Among these are charged partner $\PX^{\pm}$ states  
and a~C\nobreakdash-odd partner, 
referred to hereafter as $\PX^{\prime}_{\PC}$~state.

The~$\jpsi\Peta$~final state is well suited to search 
for the~hypothetical $\PX^{\prime}_{\PC}$~state.
Searches for 
the~\mbox{$\decay{\PX^{\prime}_{\PC}}{\jpsi\Peta}$}
decay have been performed
by the~\babar and \belle collaborations  using
\mbox{$\decay{\Bu}{\jpsi\Peta\Kp}$}~decays~\cite{BaBar:2004iez,
Belle:2013vio}. 
No~\mbox{$\decay{\PX^{\prime}_{\PC}}{\jpsi\Peta}$}
signal is observed, and 
upper limits at 90\% confidence level\,(CL)
on the~product of branching fractions 
for the~$\decay{\Bu}{\PX^{\prime}_{\PC}\Kp}$
and $\decay{\PX^{\prime}_{\PC}}{\jpsi\Peta}$
decays 
are set
to be 
$7.7\times 10^{-6}$\,(BaBar) 
and 
$3.8\times 10^{-6}$\,(Belle).

In this paper a~study 
of the~$\jpsi\Peta$~mass spectrum 
from~\decay{\Bp}{\jpsi\Peta\Kp} 
decays\footnote{Inclusion of charge-conjugate states 
is implied throughout the paper.}
is reported.
In particular 
searches for contributions 
from new hypothetical states, denoted 
hereafter as \PX,
or known charmonia 
or charmonium\nobreakdash-like
resonances are  performed.
The~study uses a~dataset collected with 
the~LHCb detector at 7, 8 and 13 \tev 
centre-of-mass energies corresponding
to an~integrated luminosity of 9\invfb.
The~results are reported in the~form of
a~ratio of branching fractions 
using 
the~normalisation
decay mode 
\mbox{$\decay{\Bp}{\left(\decay{\psitwos}
{\jpsi\Peta}\right)\Kp}$}, 
\begin{equation}
F_{\PX}  \equiv  
\dfrac{\BR(\decay{\Bp}{\PX\Kp}) 
\times \BR(\decay{\PX}{\jpsi\Peta})}
      {\BR(\decay{\Bp}{\psitwos\Kp}) 
      \times \BR(\decay{\psitwos}{\jpsi\Peta}) } \, ,
      \label{eq:FX} 
\end{equation}
and as the product of branching fractions
\begin{equation}
B_{\PX}  \equiv  
\BR(\decay{\Bp}{\PX\Kp}) \times \BR(\decay{\PX}{\jpsi\Peta})\, ,
     \label{eq:BX}
\end{equation}
the~latter obtained from the~$F_{\PX}$ ratio using 
the~known values 
of the~\mbox{$\decay{\Bp}{\psitwos\Kp}$} 
and \mbox{$\decay{\psitwos}{\jpsi\Peta}$}~branching
fractions~\mbox{\cite{ 
BaBar:2004iez,
Belle:2013vio,
PDG2021}.} 
The~results are obtained for
masses of the~hypothetical \PX~state
in the~region 
between the
$\jpsi\Peta$~threshold and 
4.65\gevcc. 
The~$F_{\PX}$~ratio and 
the~product of 
branching fractions $B_{\PX}$
are also measured for 
the~$\Ppsi(3770)$,
$\Ppsi_2(3823)$~\cite{LHCb-PAPER-2020-009},
$\Ppsi_3(3842)$~\cite{LHCb-PAPER-2019-005}, 
$\Ppsi(4040)$,
$\Ppsi(4160)$,
$\Ppsi(4415)$~charmonium states~\cite{PDG2021}, 
as well as 
the~charmonium\nobreakdash-like
$\PR(3760)$, 
$\PR(3790)$~\cite{Ablikim:2020jfw},
$\PZ_{\cquark}(3900)^0$~\mbox{\cite{BESIII:2015cld,Xiao:2013iha}},
$\Ppsi(4230)$~\cite{BESIII:2016bnd},
$\Ppsi(4360)$~\mbox{\cite{BaBar:2006ait,Belle:2007umv,BaBar:2012hpr}}
and 
$\Ppsi(4390)$~states~\mbox{\cite{BESIII:2016adj,BESIII:2020bgb},}
the~hypothetical neutral partner 
of the~charged 
$\PZ_{\cquark}(4430)^+$~state~\mbox{\cite{Belle:2007hrb,
Belle:2013shl,
LHCb-PAPER-2014-014,
Belle:2014nuw,
LHCb-PAPER-2015-038},} 
referred to as $\PZ_{\cquark}(4430)^0$ hereafter, 
and finally for the~$\PX^{\prime}_{\PC}$~state.

\section{Detector and simulation}
\label{sec:Detector}

The \lhcb detector~\mbox{\cite{LHCb-DP-2008-001,LHCb-DP-2014-002}} is 
a~single\nobreakdash-arm forward spectrometer 
covering the~\mbox{pseudorapidity} range \mbox{$2<\eta <5$}, 
designed for the study of particles containing \bquark~or \cquark~quarks.
The~detector includes a~high\nobreakdash-precision tracking system consisting
of a~silicon\nobreakdash-strip vertex detector surrounding 
the proton\nobreakdash-proton\,($\proton\proton$) interaction region, 
a~large\nobreakdash-area silicon\nobreakdash-strip detector 
located upstream of a~dipole magnet with a~bending power of 
about~$4{\mathrm{\,Tm}}$, and three stations of 
silicon\nobreakdash-strip detectors 
and straw drift tubes placed downstream of the~magnet.
The~tracking system provides a~measurement of the~momentum, \ptot, 
of charged particles with a~relative uncertainty 
that varies from 
0.5\% at low momentum to 1.0\% at~200\gevc.
The~minimum distance of a~track to a~primary 
$\proton\proton$ collision vertex\,(PV), 
the~impact parameter, 
is measured with a~resolution of~$(15+29/\pt)\mum$, 
where \pt~is the~component of the~momentum transverse 
to the~beam, in\,\gevc.
Different types of charged hadrons 
are distinguished using information
from two ring\nobreakdash-imaging Cherenkov detectors.
Photons, electrons and hadrons are identified by 
a~calorimeter system consisting of scintillating\nobreakdash-pad 
and preshower detectors, an~electromagnetic
and a~hadronic calorimeter~\cite{LHCb-DP-2020-001}.
Muons are identified by a~system composed of alternating layers of iron and 
multiwire proportional chambers.

The~online event selection is performed by a~trigger, 
which consists of a~hardware stage, based on information 
from the~calorimeter and muon systems, 
followed by a~software stage, which applies a~full event reconstruction.
At~the~hardware trigger stage, events are required to have 
a~muon with high transverse momentum or dimuon candidates 
in which the~product of the~\pt of the~muons 
has a~high value. 
In~the~software trigger, two oppositely charged muons are required 
to form a~good\nobreakdash-quality vertex that is significantly 
displaced from every~PV, with a~dimuon mass exceeding $2.7\gevcc$.

Simulated events are used to describe signal shapes 
and to compute the~efficiencies needed to determine 
the~branching fraction ratios.
In~the~simulation, $\proton\proton$~collisions are 
generated using \pythia~\cite{Sjostrand:2007gs} with 
a~specific \lhcb configuration~\cite{LHCb-PROC-2010-056}.
Decays of unstable particles are 
described by \evtgen~\cite{Lange:2001uf}, 
in which final\nobreakdash-state radiation is generated 
using \photos~\cite{
davidson2015photos}.
The~interaction of the~generated particles with the~detector, 
and its response, are implemented using 
the~\geant toolkit~\cite{Allison:2006ve, *Agostinelli:2002hh} 
as described in Ref.~\cite{LHCb-PROC-2011-006}.
The~transverse momentum and rapidity, $y$, 
spectra of the~$\Bu$~mesons in simulation 
are corrected to represent better 
those observed in data. 
The~correction factors are calculated 
by comparing the~observed
$\pt$ and $y$~spectra
for a~high\nobreakdash-yield and 
low\nobreakdash-background sample of 
reconstructed \decay{\Bu}{\jpsi\Kp}~decays
with corresponding simulated samples. 
In~the~simulation, 
the~\mbox{$\decay{\Bu}{\jpsi\Peta\Kp}$}~decays 
are produced according  to a~phase-space decay model. 
Simulated decays are corrected to reproduce 
the~$\jpsi\Peta$ and $\Peta\Kp$~mass distributions 
observed in data. 
To~describe accurately the~variables used 
for kaon identification,  
the~corresponding  quantities in simulation 
are resampled according to values obtained from
calibration data samples of 
\mbox{$\decay{\Dstarp}{\left( \decay{\Dz}{\Km\pip}\right)\pip}$}~decays~\cite{LHCb-DP-2018-001}. 
The~procedure accounts for correlations between 
the~variables associated to a~particular track, 
as well as the~dependence of 
the~kaon identification response
on $\pt$, $\eta$~and  the~multiplicity of tracks in the event. 
To~account for imperfections in the~simulation of 
charged\nobreakdash-particle reconstruction, 
the~track reconstruction efficiency determined from simulation
is corrected using control channels in data~\cite{LHCb-DP-2013-002}.

\section{Event selection}
Candidate \decay{\Bp}{\jpsi\Peta\Kp} decays 
are reconstructed through 
the~\decay{\jpsi}{\mumu} and \decay{\Peta}{\g\g} decay modes. 
A~loose initial selection 
is applied to reduce 
the~background.
The~criteria are chosen to be similar 
to those used in 
previous \lhcb studies~\mbox{\cite{LHCb-PAPER-2012-022,
LHCb-PAPER-2012-053,
LHCb-PAPER-2013-024,
LHCb-PAPER-2014-008,
LHCb-PAPER-2014-056,
LHCb-PAPER-2021-003}}.
Subsequently, a~multivariate estimator based 
on an~artificial neural 
network algorithm~\mbox{\cite{McCulloch,rosenblatt58}}, 
configured with 
a~cross-entropy cost estimator~\cite{Zhong:2011xm}, 
in the~following referred 
to as the~{\sc{MLP}}~classifier, is applied.

Muon and kaon candidates are identified by combining
information from the~Cherenkov~detectors, 
calorimeters and muon detectors~\cite{LHCb-PROC-2011-008}
associated to the~reconstructed tracks.
Transverse momenta of muon candidates are 
required to be greater than 550\mevc. 
To~reduce combinatorial background only tracks 
that are inconsistent with originating 
from any reconstructed PV in the~event are considered. 
Pairs of oppositely charged muons consistent 
with originating from a~common vertex 
are combined to form \decay{\jpsi}{\mumu}~candidates. 
The~reconstructed mass of the~pair is required to be
between  $3.056$~and $3.136\gevcc$.

Photons are reconstructed from clusters 
in the~electromagnetic calorimeter
that have transverse energy larger than 500\mev
and are not associated with 
reconstructed tracks~\mbox{\cite{Terrier:691743,
LHCb-DP-2020-001}}. 
Photon identification is based on the~combined information 
from electromagnetic and hadronic calorimeters,
scintillation pad and preshower detectors 
and the~tracking system.
Candidate \decay{\Peta}{\g\g} decays are
reconstructed as diphoton combinations with 
mass within 
$\pm 60$~\mevcc
of the~known 
$\Peta$ mass~\cite{PDG2021} 
and transverse momentum greater than 1.5~\gevc.

The selected $\jpsi$~candidates are combined with $\Kp$~and $\Peta$~candidates
to form $\Bp$~candidates. 
Each~$\Bp$~candidate is associated with
 the~PV that yields the~smallest~$\chisqip$, 
 where \chisqip is defined 
 as the~difference in the~vertex\nobreakdash-fit 
 \chisq of a~given PV 
 reconstructed with and without 
the~charged tracks that form
the~\Bu~candidate under consideration.
 To~improve
 the~\Bp~mass resolution 
 a~kinematic fit~\cite{Hulsbergen:2005pu}  is performed. 
 This~fit constrains
 the~$\mumu$ and $\g\g$~masses 
 to the~known
 $\jpsi$ and $\Peta$~mesons masses~\cite{PDG2021}, respectively,
 and the~\Bu~candidate 
 to originate from its associated PV. \
The~proper decay time of the~\Bp~candidate 
is required  to be greater than $200\mum/c$ to suppress the
large combinatorial background. 

A~further selection based on the~{\sc{MLP}}~classifier 
reduces the combinatorial background 
to a~low level whilst retaining a~high signal efficiency.
Variables included in the classifier are related 
to the~reconstruction quality, kinematics and decay time of 
the~$\Bp$~candidates, kinematics of 
the~final\nobreakdash-state particles
and a~variable that characterises kaon identification.
The~classifier
is trained using simulated samples 
of \decay{\Bp}{\jpsi\Peta\Kp} decays as signal proxy.
The~\mbox{\decay{\Bp}{\jpsi\Peta\Kp}}~candidates from 
data with mass,  $m_{\jpsi\Peta\Kp}$, 
ranging between 5.4 and 5.7\gevcc, are 
used as background proxy.
To~avoid introducing a~bias in the~{\sc{MLP}}~evaluation,
a~$k$-fold cross-validation
technique~\cite{geisser1993predictive} 
with $k=13$ is used.

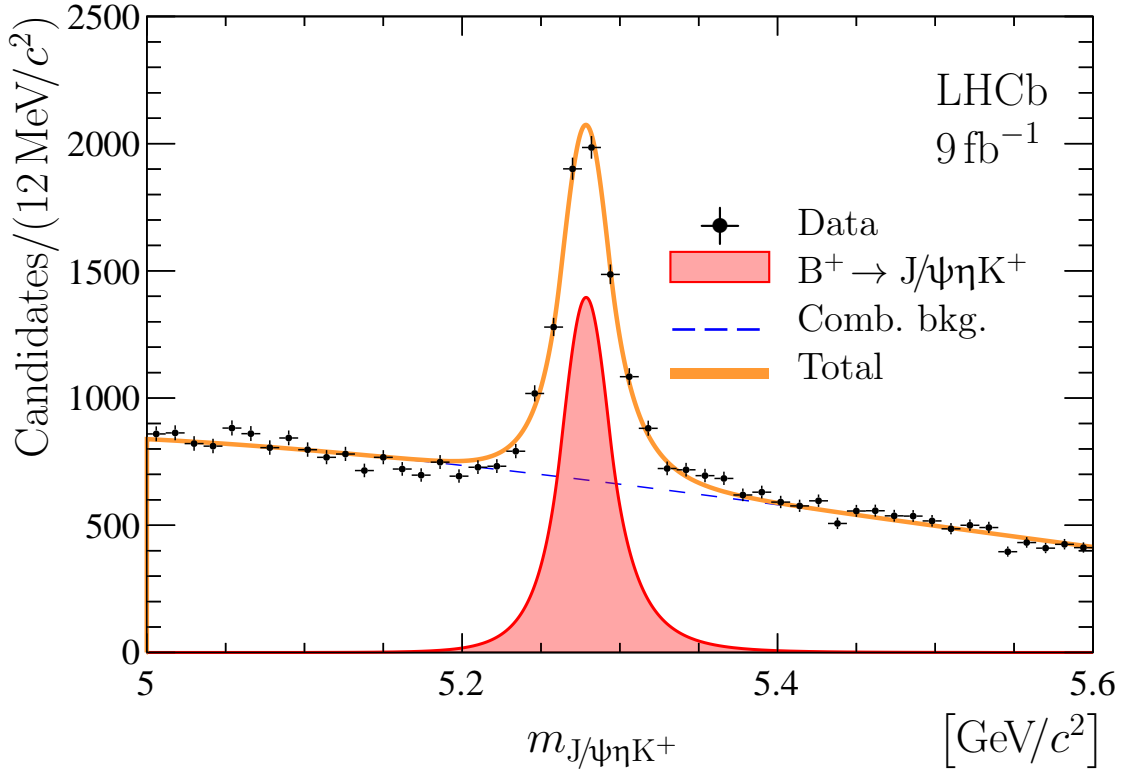
\begin{figure}[t]
	\setlength{\unitlength}{1mm}
	\centering
	\begin{picture}(150,105)
 	\put(0,2){\includegraphics*[width=150mm]{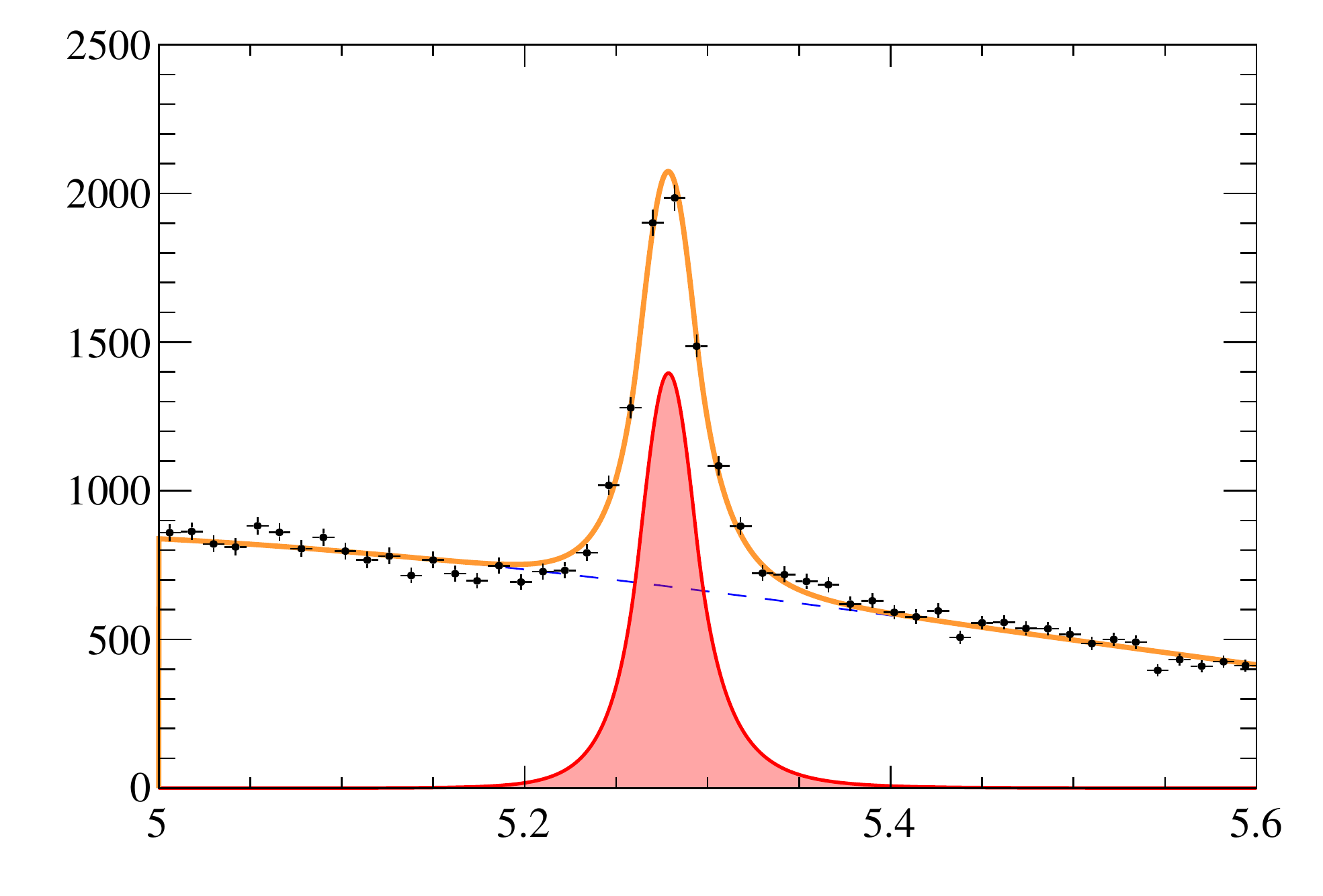}}
	\put( 68, 1){\Large$m_{\jpsi\Peta\Kp}$}
    \put(122, 1){\Large$\left[\!\gevcc\right]$}
	 \put( 0,40){\Large\begin{sideways}{Candidates/$(12\mevcc)$}\end{sideways}}
	\put(120,83){\Large$\begin{array}{l}\lhcb \\ 9\invfb\end{array}$} 
    \put(85,60){\large$\begin{array}{cl}
    \bigplus\mkern-18mu\bullet\ \  & \mathrm{Data} 
    \\ 
    {\begin{tikzpicture}[x=1mm,y=1mm]\filldraw[fill=red!35!white,draw=red,thick] (0,0) rectangle (13,4);\end{tikzpicture}}  & \decay{\Bp}{\jpsi\Peta\Kp} 
    \\
    {\color{blue}     {\hdashrule[0.0ex][x]{13mm}{1.0pt}{3.0mm 1.0mm}}}
    &  \mathrm{Comb.\ bkg.}
    \\
    {\color[RGB]{255,153,51} {\rule{13mm}{4.0pt}}}
    &  \mathrm{Total}
    \end{array}$}
	\end{picture}
	\caption {Mass distribution of  selected 
	\mbox{$\decay{\Bp}{\jpsi\Peta\Kp}$}~candidates. 
	The~result of the~fit, described 
	in the~text, is overlaid.
	}
	\label{fig:datavsmc}
\end{figure}

The~requirement on the~{\sc{MLP}} classifier 
is chosen to maximize 
the~figure\nobreakdash-of\nobreakdash-merit defined as 
\mbox{$S/\sqrt{B+S}$}, 
where $S$~represents 
the~expected signal yield from simulation, 
and $B$~stands for the~background yield obtained 
by fitting the data.  
The~expected signal yield is estimated 
as $S=\varepsilon S_0$, where 
$\varepsilon$ is the~efficiency 
of the~requirement on 
the~{\sc{MLP}}~classifier determined from simulation,
and $S_0$~is the~signal yield 
obtained from the~fit to 
the~data when no 
requirement 
is applied. 
The~mass  distribution of  selected 
\mbox{$\decay{\Bp}{\jpsi\Peta\Kp}$}~candidates 
is shown in Fig.~\ref{fig:datavsmc}.

\section{ $\decay{\Bp}{\jpsi\Peta\Kp}$~signal
and $\jpsi\Peta$~mass spectrum}  
\label{sec:signals}

The~\mbox{$\decay{\Bu}{\jpsi\Peta\Kp}$}~signal 
yield is determined using 
an~extended 
unbinned maximum\nobreakdash-likelihood fit to 
the~$\jpsi\Peta\Kp$~mass 
distribution with signal and background components.
The~signal 
is modelled by a~modified Gaussian function 
with power\nobreakdash-law tails on
both sides of 
the~distribution~\mbox{\cite{Skwarnicki:1986xj,
LHCb-PAPER-2011-013},}
referred to hereafter as $\mathcal{F}_{\mathcal{S}}$.  
The~tail parameters of the~modified Gaussian function
are fixed from simulation, 
while the~peak position and resolution 
are allowed to vary in the~fit. 
The~combinatorial background
is parametrised with an~exponential function.
The~fit result is overlaid in Fig.~\ref{fig:datavsmc}
and the~signal yield 
is found to be 
\begin{equation*}
    N_{\decay{\Bp}{\jpsi\Peta\Kp}} = \left(5.39\pm0.16\right)\times 10^{3}\,.
\end{equation*}

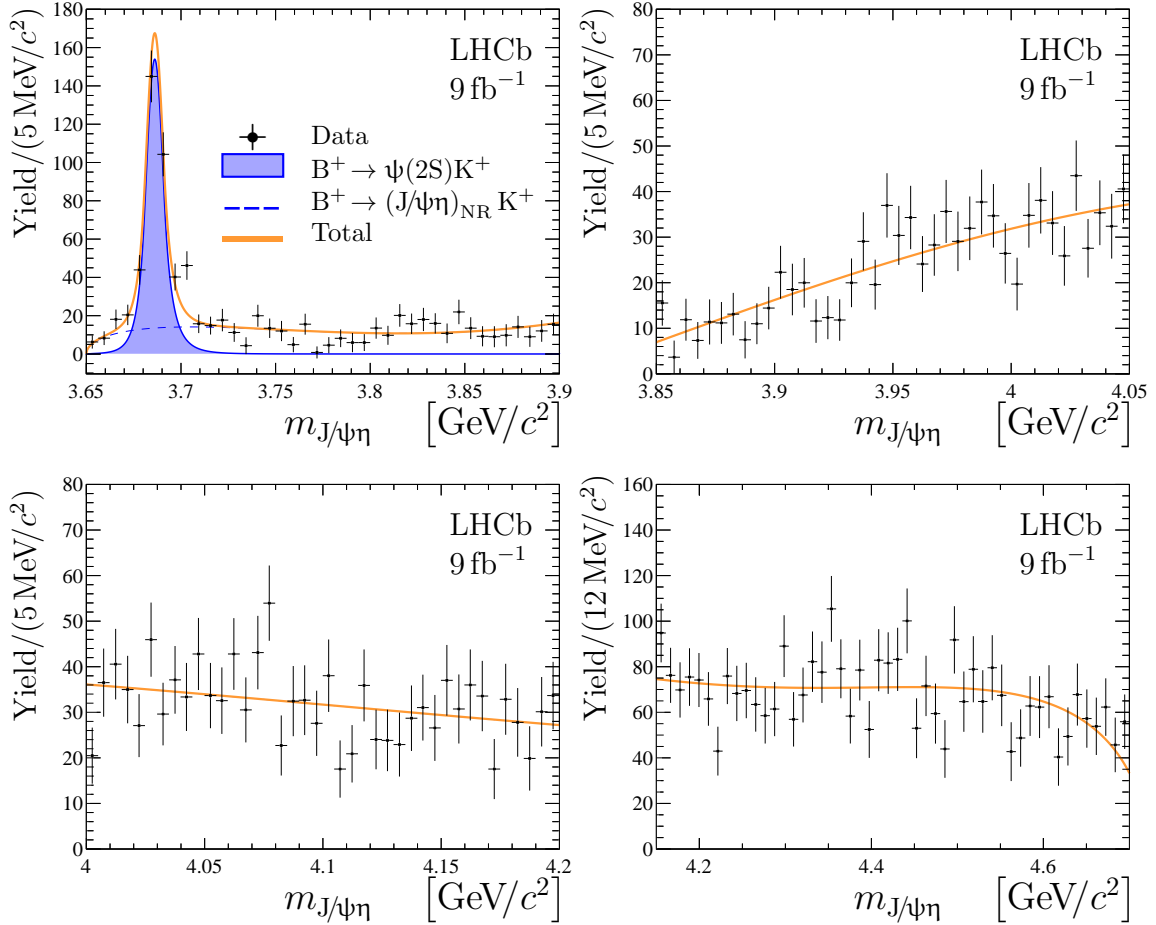
\begin{figure}[t]
  \setlength{\unitlength}{1mm}
  \centering
  \begin{picture}(150,123)
    \definecolor{root8}{rgb}{0.35, 0.83, 0.33}
    
    \put( 0,64){\includegraphics*[width=75mm]{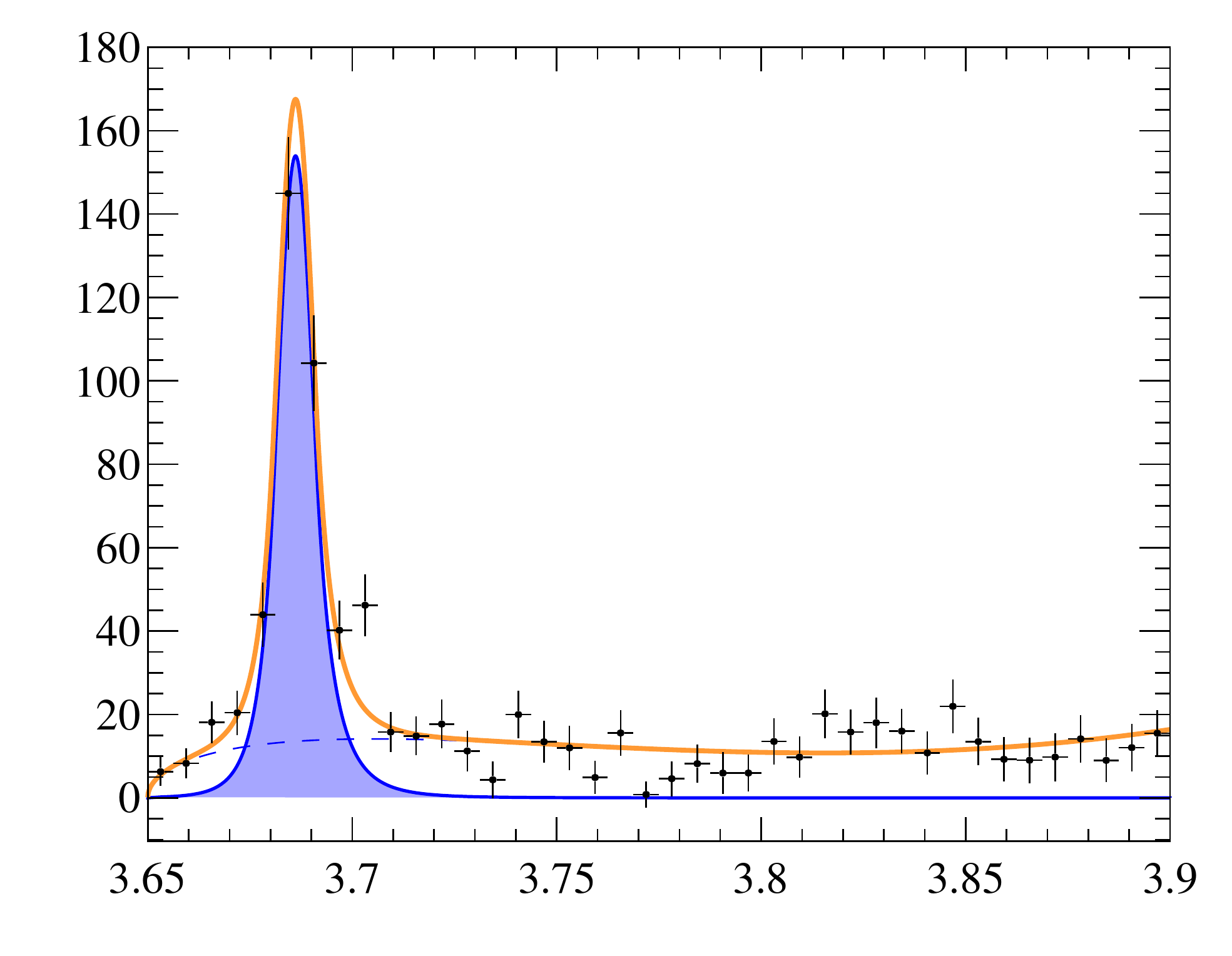}}
    \put(75,64){\includegraphics*[width=75mm]{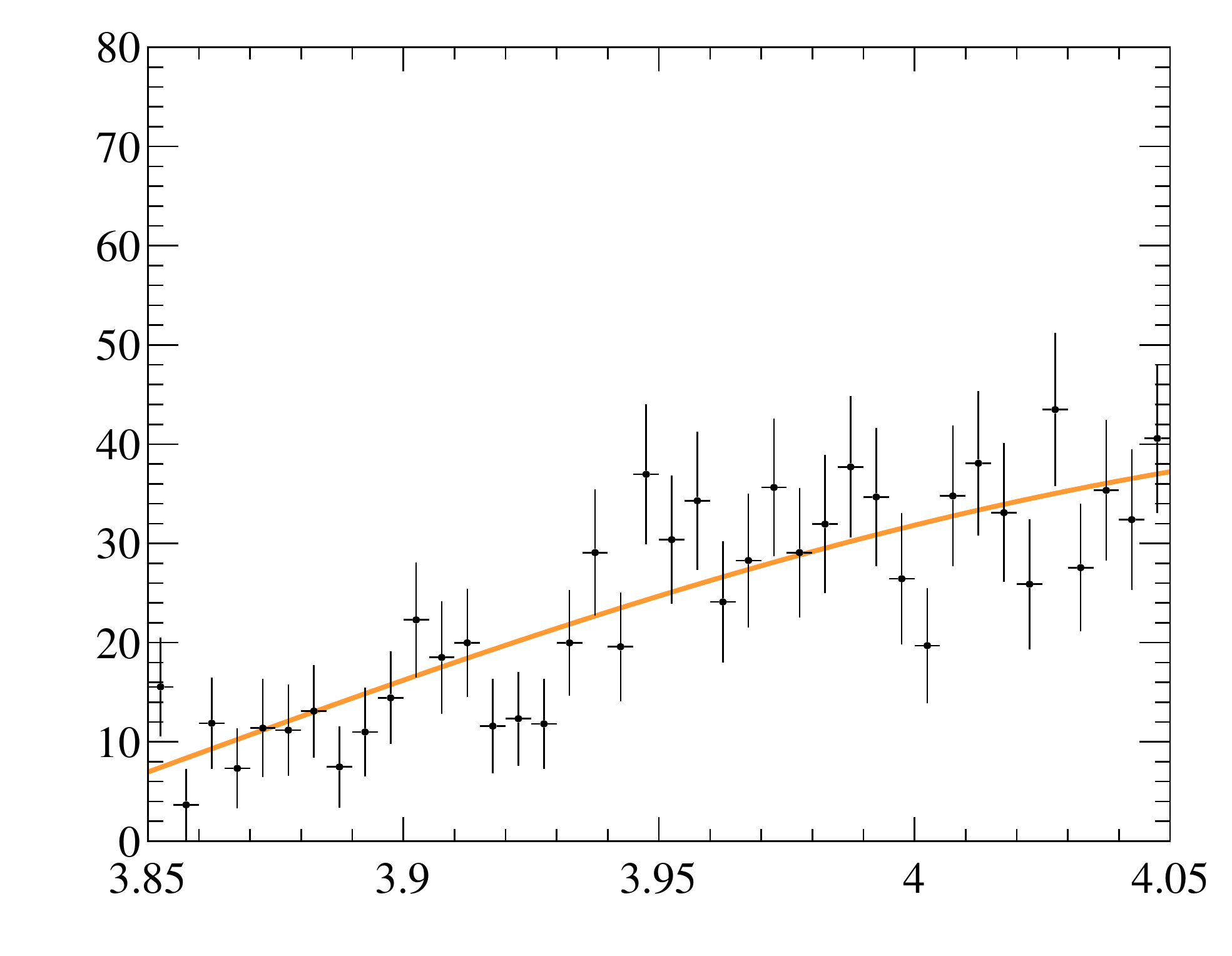}}
    \put( 0, 1){\includegraphics*[width=75mm]{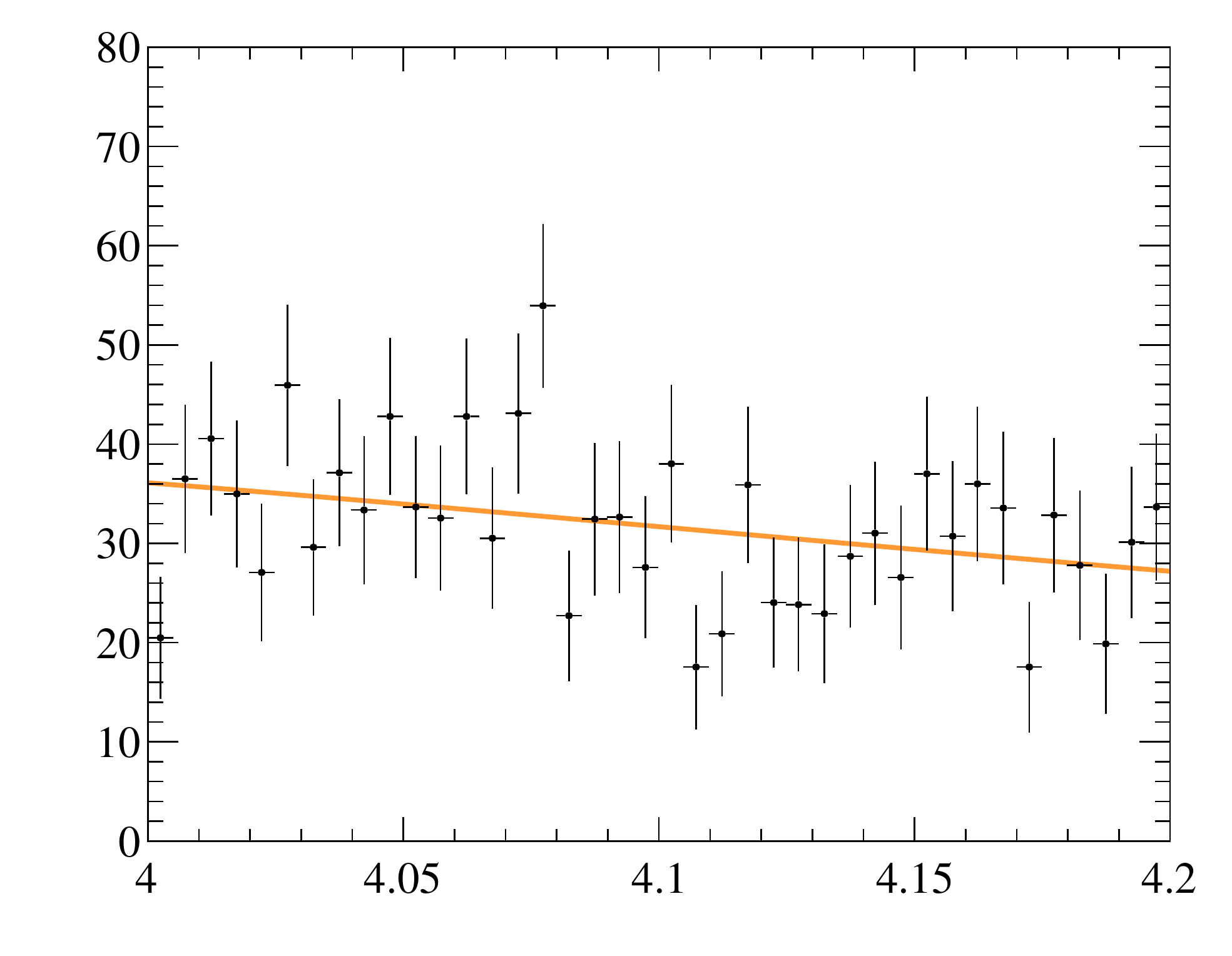}}
    \put(75, 1){\includegraphics*[width=75mm]{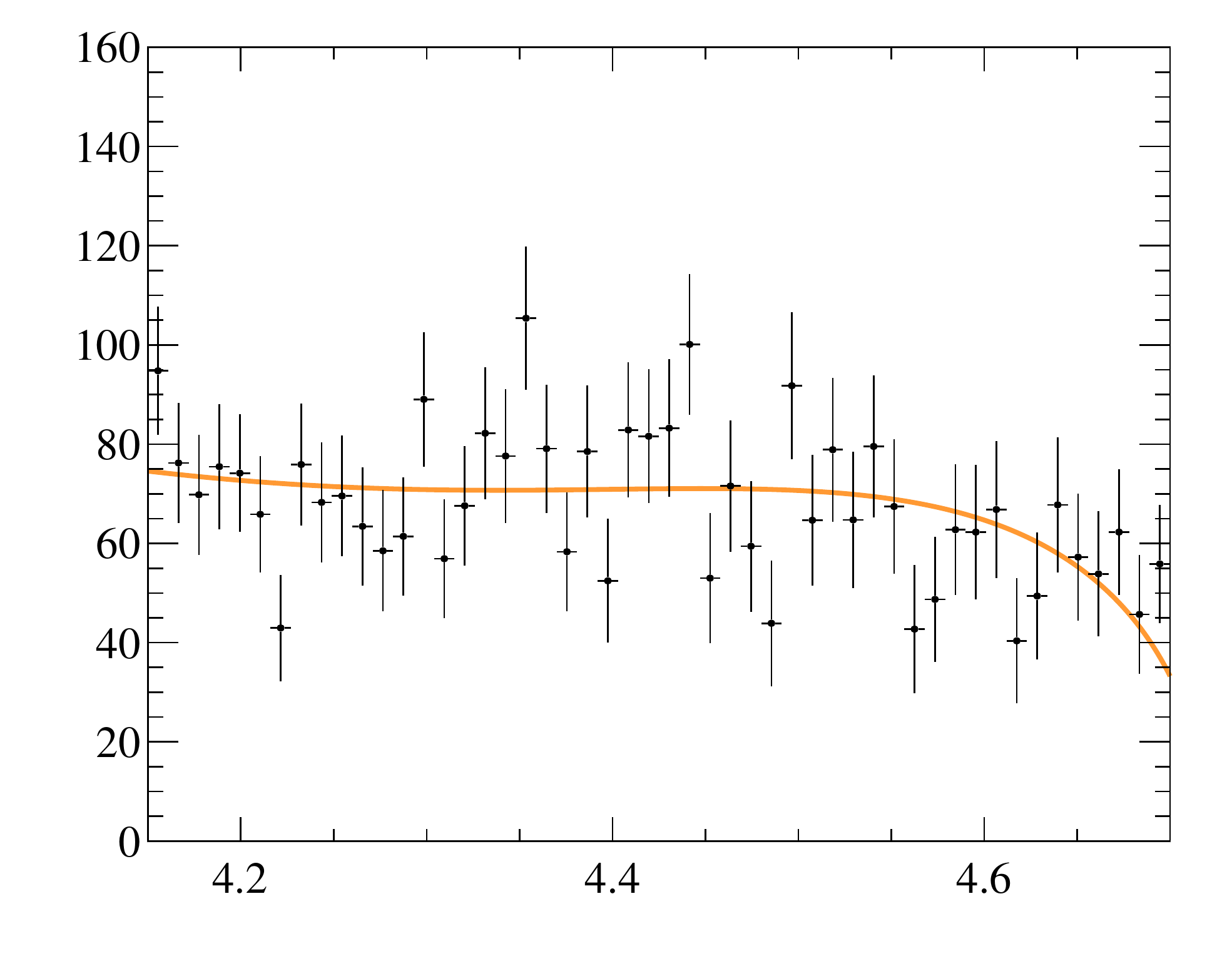}}

    \put(  -1,89){\begin{sideways}{Yield/$(5\mevcc)$}\end{sideways}}
    \put(  -1,25){\begin{sideways}{Yield/$(5\mevcc)$}\end{sideways}}
    \put(  74,89){\begin{sideways}{Yield/$(5\mevcc)$}\end{sideways}}
    \put(  74,25){\begin{sideways}{Yield/$(12\mevcc)$}\end{sideways}}

    \put(35  , 0){\large$m_{\jpsi \Peta}$}
    \put(35  ,63){\large$m_{\jpsi \Peta}$}
    \put(110 , 0){\large$m_{\jpsi \Peta}$}
    \put(110 ,63){\large$m_{\jpsi \Peta}$}

    \put(53 , 0){\large$\left[\!\gevcc\right]$}
    \put(53 ,63){\large$\left[\!\gevcc\right]$}
    \put(128, 0){\large$\left[\!\gevcc\right]$}
    \put(128,63){\large$\left[\!\gevcc\right]$}
  
    \put( 55,110){$\begin{array}{l}\lhcb \\ 9\invfb\end{array}$}
    \put( 55, 47){$\begin{array}{l}\lhcb \\ 9\invfb\end{array}$}
    \put(130,110){$\begin{array}{l}\lhcb \\ 9\invfb\end{array}$}
    \put(130, 47){$\begin{array}{l}\lhcb \\ 9\invfb\end{array}$}
 
     \put(25,95){\footnotesize$\begin{array}{cl}	 
     \bigplus\mkern-18mu\bullet\ \  & \mathrm{Data} 
     \\
     \begin{tikzpicture}[x=1mm,y=1mm]\filldraw[fill=blue!35!white,draw=blue,thick]  (0,0) rectangle (8,3);\end{tikzpicture} 
      & \decay{\Bp}{\psitwos\Kp}
      \\
     {\color{blue}{\hdashrule[0.0ex][x]{8mm}{1.0pt}{2.0mm 0.3mm}}}
      &  \decay{\Bp}{\left(\jpsi\Peta\right)_{\mathrm{NR}}\Kp}
      \\
      {\color[RGB]{255,153,51} {\rule{8mm}{2.0pt}}}
      & \mathrm{Total}
     \end{array}$}
 
  \end{picture}
  \caption {\small 
    Background-subtracted $\jpsi\Peta$~mass distribution
	from~\mbox{$\decay{\Bp}{\jpsi\Peta\Kp}$}~decays
	in four 
	$\jpsi\Peta$~mass regions.
	The~results of the~fits
	without contributions 
	from a narrow \PX~state, described 
	in the~text, are overlaid.}
  \label{fig:ranges}
\end{figure}

The~search for 
the~\mbox{$\decay{\Bp}
{ \left( \decay{\PX}{\jpsi\Peta}\right)\Kp}$}~signal 
is performed using 
extended unbinned 
maximum\nobreakdash-likelihood 
fits to 
the~background\nobreakdash-subtracted 
$\jpsi\Peta$~mass spectrum.
The~$sPlot$~technique~\cite{Pivk:2004ty}
is used for 
the~background subtraction 
using the~$\jpsi\Peta\Kp$~mass 
as the~discriminating variable.
To~improve 
the~$\jpsi\Peta$~mass resolution
and to eliminate a~small correlation
between the $m_{\jpsi\Peta\Kp}$ and 
$m_{\jpsi\Peta}$~variables,
the~$\jpsi\Peta$~mass is computed 
using a~kinematic 
fit~\cite{Hulsbergen:2005pu}
that constrains the~mass of the~\Bp~candidate
to its known value~\cite{PDG2021}.
For~easier parametrisation of 
the~nonresonant 
component,  
the~fit to 
the~$\jpsi\Peta$~mass distribution
is performed separately in~four different 
overlapping 
mass regions. 
For~\PX~masses below $3.875\gevcc$,
a~fit of 
the~lowest-mass region, 
\mbox{$3.65<m_{\jpsi\Peta}<3.90\gevcc$},
is performed. 
The~$\jpsi\Peta$~mass regions 
\mbox{$3.85<m_{\jpsi\Peta}<4.05\gevcc$},
\mbox{$4.0<m_{\jpsi\Peta}<4.2\gevcc$},
and 
\mbox{$m_{\jpsi\Peta}>4.15\gevcc$}
are used for 
$\PX$~masses within the~ranges 
\mbox{$3.875<m_{\PX}<4.025\gevcc$},
\mbox{$4.025<m_{\PX}<4.175\gevcc$},
and 
\mbox{$m_{\PX}>4.175\gevcc$}, 
respectively. 
The~background-subtracted $\jpsi\Peta$~mass 
distribution 
for these regions is shown
in~Fig.~\ref{fig:ranges}. 
A~clear narrow peak,  
corresponding to 
the~\mbox{$\decay{\Bp}{\left(\decay{\psitwos}
{\jpsi\Peta}\right)\Kp}$}~decay,
is visible in the~low\nobreakdash-mass region. 
This signal is used as a~normalisation channel. 

The~fit function 
to the~lowest\nobreakdash-mass region 
consists of three components:
\begin{enumerate}
    \item the decay of interest $\decay{\Bp}{\left(\decay{\PX}{\jpsi\Peta}\right)\Kp}$, 
    referred to as $\mathcal{C}_{\PX}$~component; 
    \item the~$\decay{\Bp}{\left(\decay{\psitwos}
    {\jpsi\Peta}\right)\Kp}$~signal,
    referred to as $\mathcal{C}_{\psitwos}$~component; 
   \item the~\mbox{$\decay{\Bp}
   {\left(\jpsi\Peta\right)_{\mathrm{NR}}\Kp}$}~decays
   with no resonances in the~$\jpsi\Peta$~system, 
   and referred to as $\mathcal{C}_{\mathrm{NR}}$~component.
\end{enumerate}
The~$\mathcal{C}_{\psitwos}$~component 
and the~$\mathcal{C}_{\PX}$ contribution 
for $\PX$~states with the~natural width
negligible with respect to 
the~detector resolution (referred as narrow)
are modelled
using the~$\mathcal{F}_{\mathcal{S}}$~shape.
%
The~tail parameters of all 
the~$\mathcal{F}_{\mathcal{S}}$~functions are fixed 
from simulation, 
while the~peak position and resolution 
parameter for the~$\mathcal{C}_{\psitwos}$~component 
are allowed to vary 
in the~fit. The~ratio of
the~resolution parameters 
for the~$\mathcal{C}_{\PX}$ 
and $\mathcal{C}_{\psitwos}$~components
is fixed at the~value obtained from simulation.
This procedure also accounts 
for a~small imperfection in the~modelling of  
the~$\jpsi\Peta$~mass resolution 
in the~simulation~\mbox{\cite{LHCb-PAPER-2020-008,
LHCb-PAPER-2020-009,
LHCb-PAPER-2020-035}}.
The~nonresonant component  
${\mathcal{C}}_{\mathrm{NR}}$
is parameterised with a~product of 
the~phase-space function describing
a~two\nobreakdash-body system 
out of the~three\nobreakdash-body final state~\cite{Byckling}
and a~positive first\nobreakdash-order 
polynomial function.

For~\PX~masses above 3.875\gevcc,
the~fit is performed simultaneously  
in two $\jpsi\Peta$~mass regions,
one containing the~$\PX$~mass
and the~other the~\psitwos~state.
For~the~\PX~mass region 
the~fit function consists of a~pair of 
components, $\mathcal{C}_{\PX}$ and 
$\mathcal{C}_{\mathrm{NR}}$,
while the~$\psitwos$~region fit 
includes the~$\mathcal{C}_{\psitwos}$ and 
$\mathcal{C}_{\mathrm{NR}}$ components.
The~parameters for two~$\mathcal{C}_{\mathrm{NR}}$
components are independent in the~two regions.

No significant signal is found for 
\mbox{$\decay{\Bp}{\left(\decay{\PX}{\jpsi\Peta}\right)\Kp}$}~decays 
occurring via a~hypothetical narrow \PX~particle in the~\mbox{$3.7<m_{\PX}<4.7\gevcc$} mass region.
Fit results without the~$\mathcal{C}_{\PX}$~component are
shown in  Fig.~\ref{fig:ranges}, 
illustrating that  
no sizeable contribution from 
decays with a~narrow intermediate $\PX$~state 
is required to describe the~data.
To~quantify the absence of 
the~\mbox{$\decay{\Bp}
{ \left( \decay{\PX}{\jpsi\Peta} \right)\Kp }$}~signal, 
fits are performed with 
the~mass $m_{\PX}$ of the hypothetical 
\PX~particle fixed to a~value
that is scanned across the~whole available 
$\jpsi\Peta$~mass range.
The~yield of the~$\mathcal{C}_{\PX}$~component, 
$N_{\PX}$, 
is parametrised using the~yield of 
the~$\mathcal{C}_{\psitwos}$~component, $N_{\psitwos}$, 
and the~ratio of branching fractions, $F_{\PX}$,
defined by Eq.~\eqref{eq:FX}, as
\begin{subequations}\label{eq:NX}
\begin{equation}\label{eq:NXa}
  N_{\PX}(m_{\PX}) =   
  N_{\psitwos} F_{\PX} ( m_{\PX} ) 
  R_{\varepsilon} (m_{\PX}) \,, 
\end{equation}
where $R_{\varepsilon}$ 
is the~ratio of 
total efficiencies 
for 
the~\mbox{$\decay{\Bp}{\left( \decay{\PX}
{\jpsi\Peta}  \right)\Kp} $}
and \mbox{$\decay{\Bp}{\left( \decay{\psitwos}
{\jpsi\Peta}  \right)\Kp} $}~channels,
described in Sec.~\ref{sec:eff}. 
The~parameters $N_{\psitwos}$ and $F_{\PX}$ 
are left to vary  in the~fit, 
and the~uncertainty on the~ratio 
$R_{\epsilon}$ is 
included in the~fit through a~Gaussian constraint.
A~second set of fits exploits 
an~alternative parametrisation that allows 
for the~determination of the~product of the~branching fractions
$B_{\PX}$, defined by Eq.~\eqref{eq:BX}, through 
\begin{equation}
  N_{\PX}(m_{\PX})  =   
  \frac{N_{\psitwos} B_{\PX}(m_{\PX}) 
  R_{\varepsilon} (m_{\PX}) }{
  \BR\left( \decay{\Bp}{\psitwos\Kp} \right)
  \BR\left( \decay{\psitwos}{\jpsi\Peta} \right) }
  \,, 
\end{equation}
\end{subequations}
where $N_{\psitwos}$ and $B_{\PX}$
are fit parameters, and uncertainties for
the~branching fractions 
$\BR\left( \decay{\Bp}{\psitwos\Kp} \right)$ and 
$\BR\left( \decay{\psitwos}{\jpsi\Peta}
\right)$~\cite{PDG2021}
are included in the~fit using Gaussian constraints. 

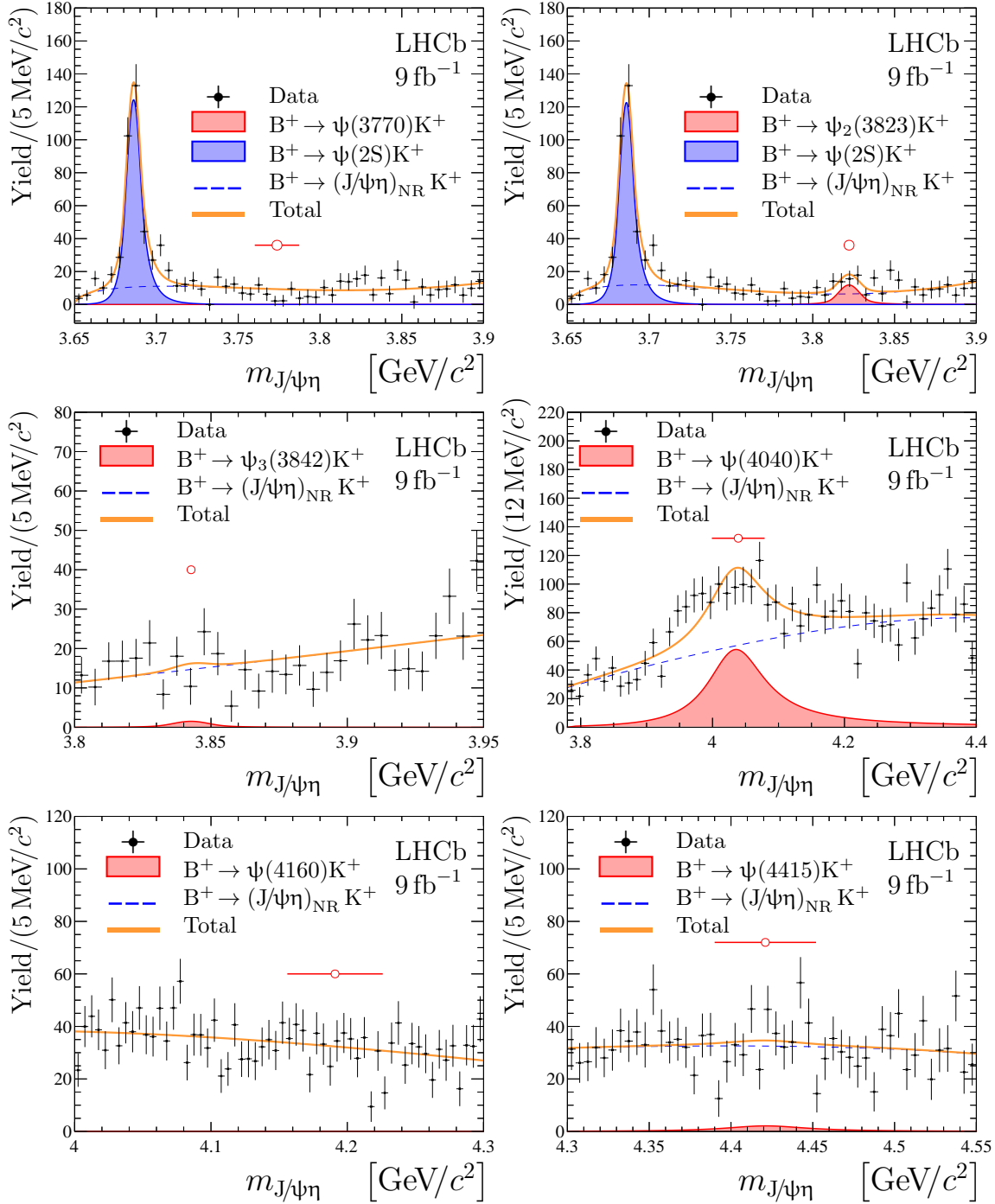
\begin{figure}[t]
  \setlength{\unitlength}{1mm}
  \centering
  \begin{picture}(150,185)
    \definecolor{root8}{rgb}{0.35, 0.83, 0.33}
    \put( 0,126){\includegraphics*[width=75mm]{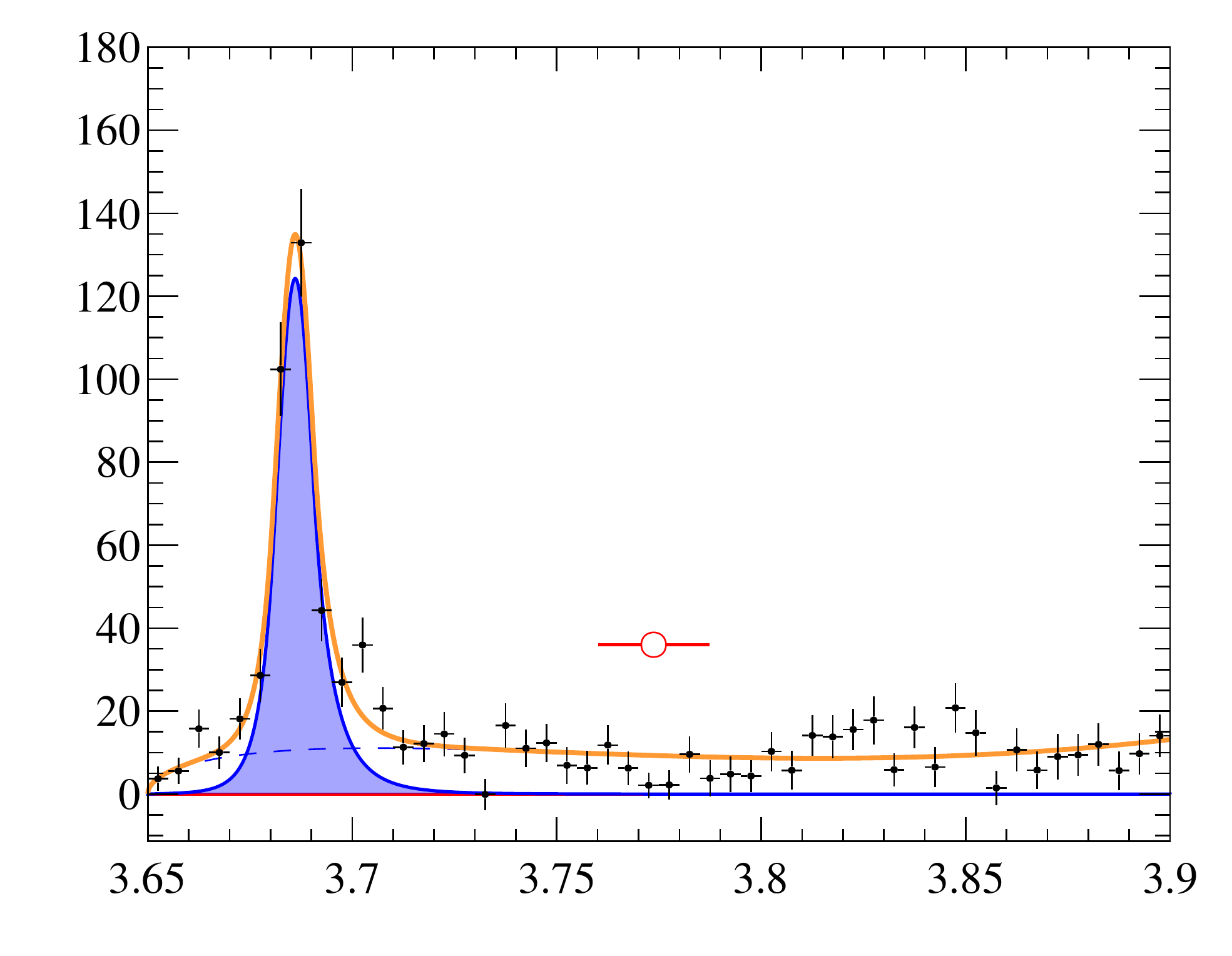}}
    \put(75,126){\includegraphics*[width=75mm]{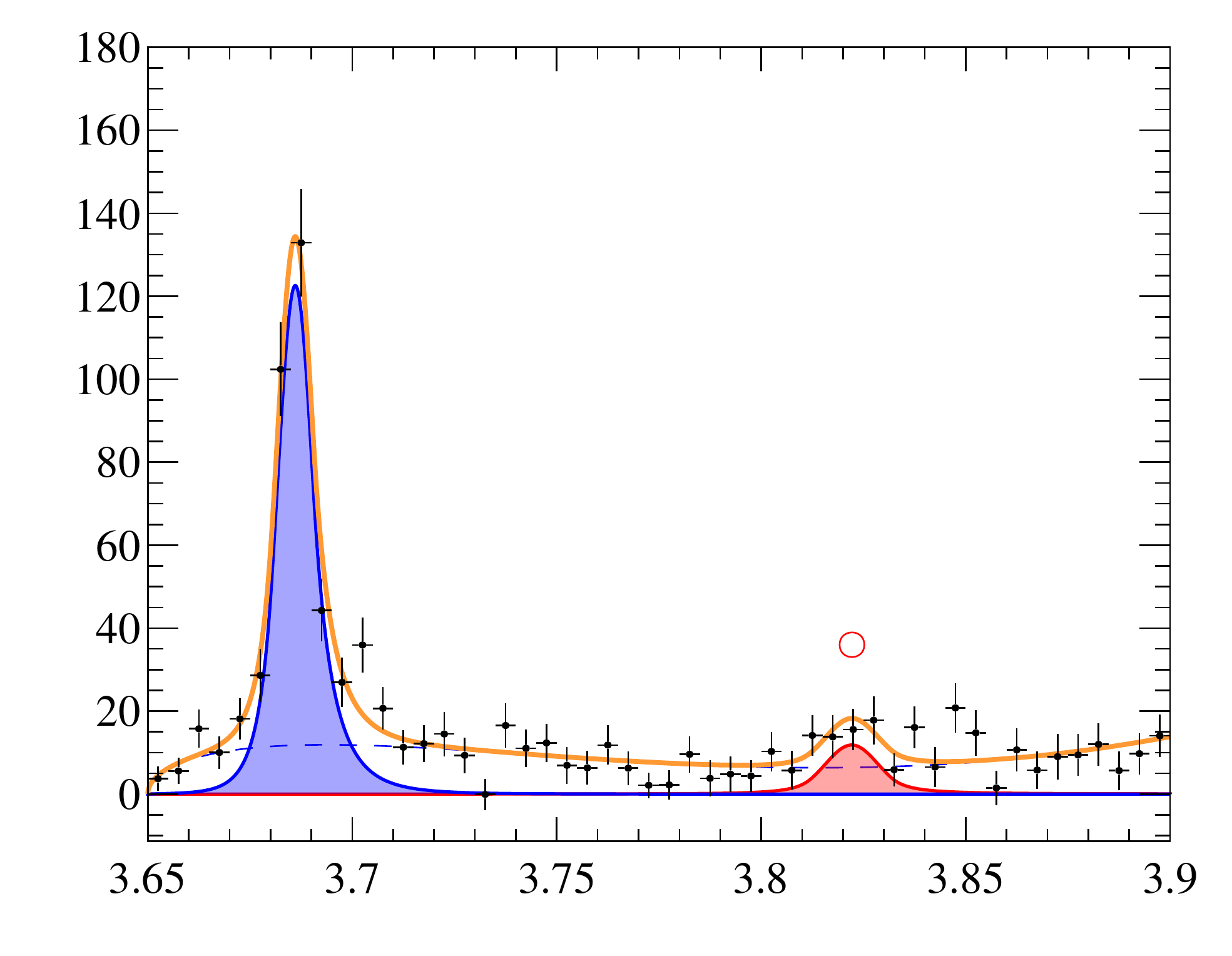}}
    \put( 0, 64){\includegraphics*[width=75mm]{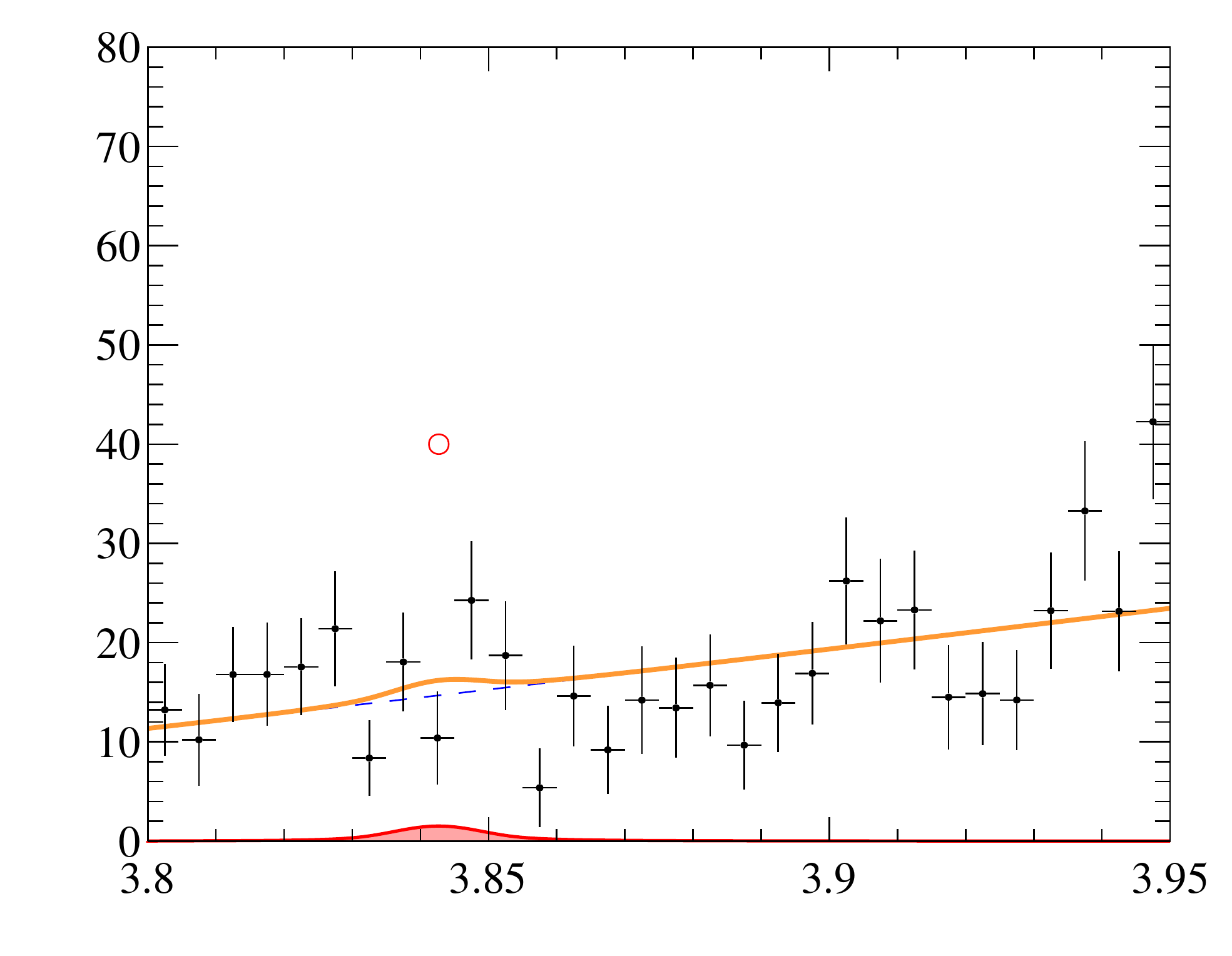}}
    \put(75, 64){\includegraphics*[width=75mm]{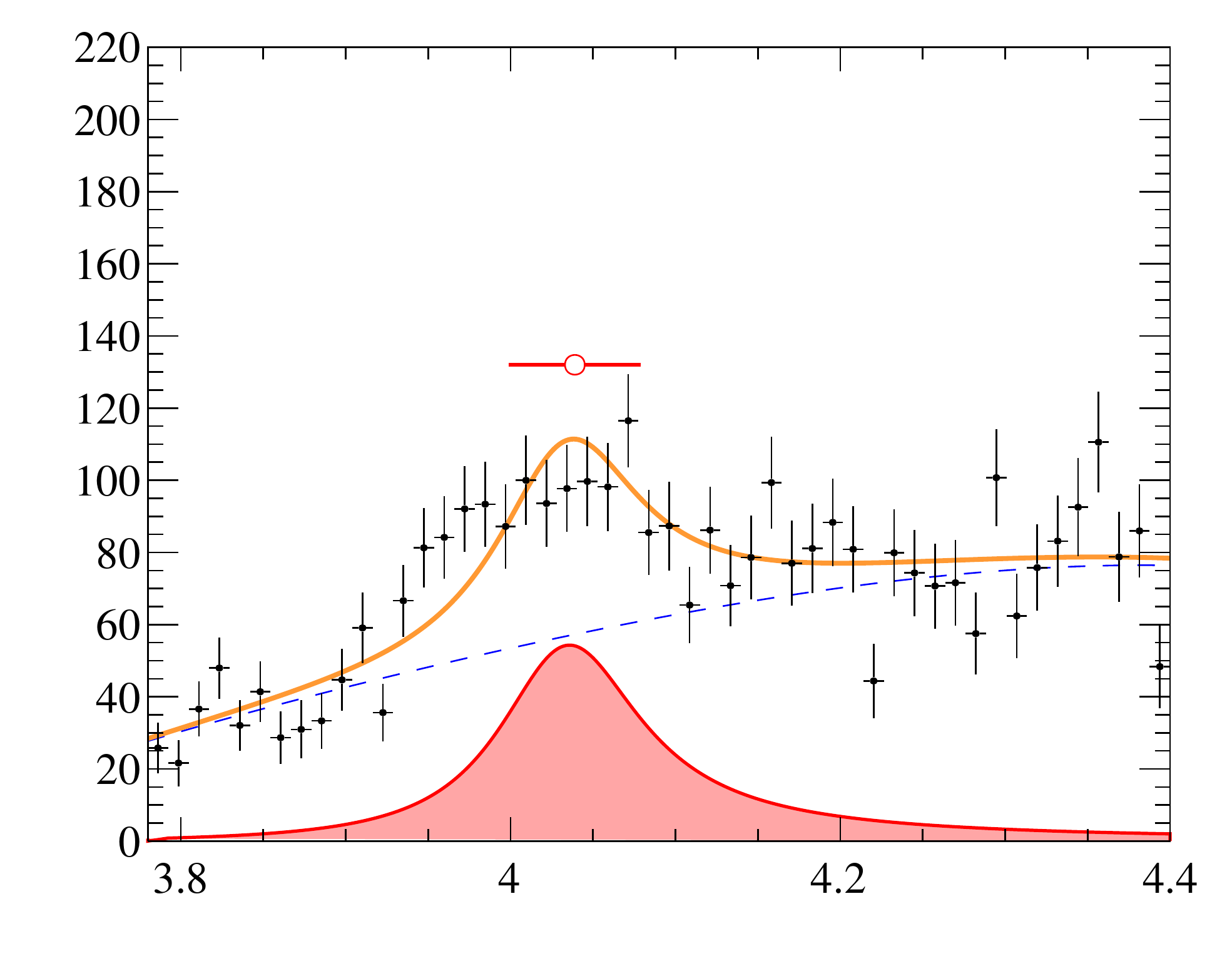}}
    \put( 0,  2){\includegraphics*[width=75mm]{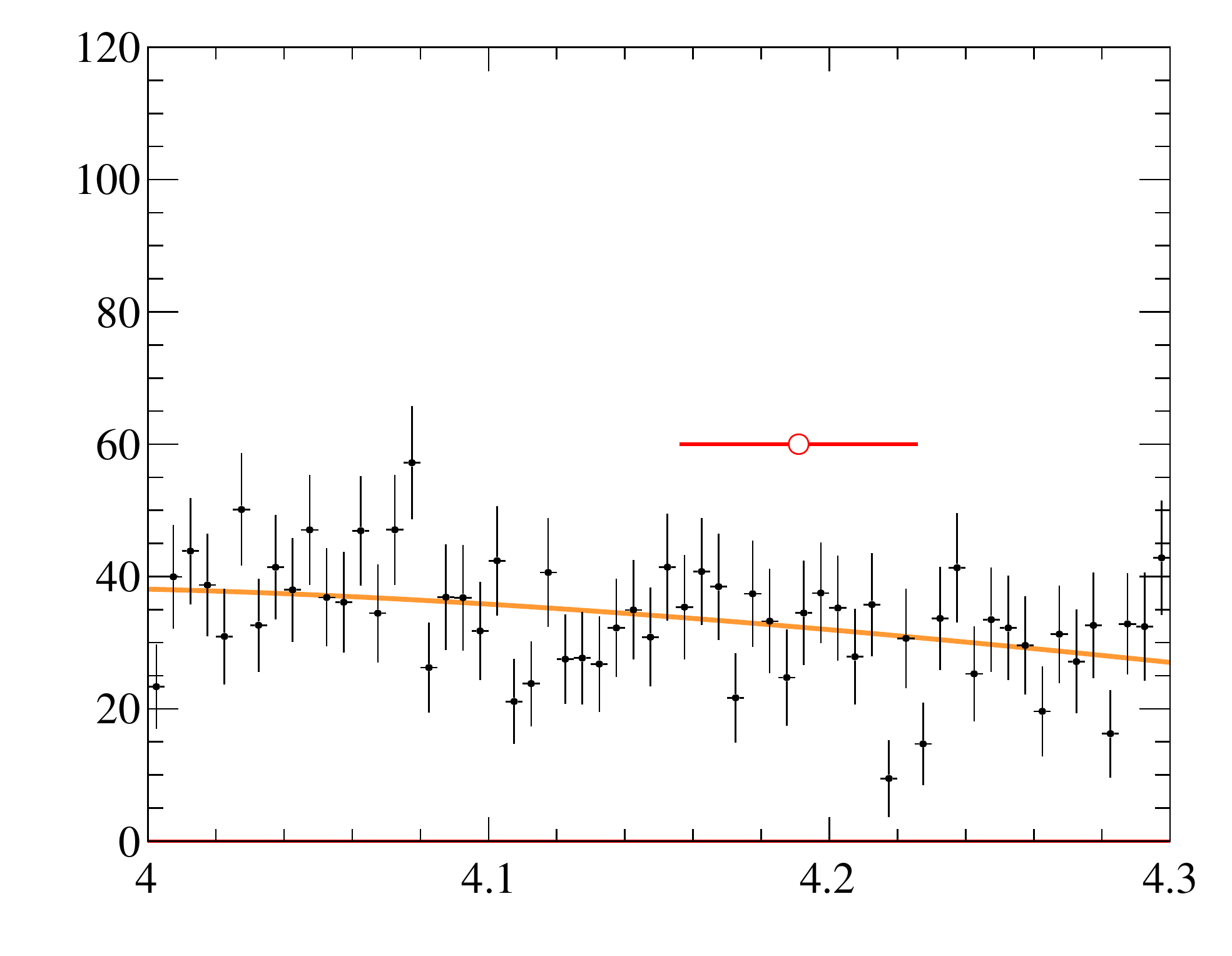}}
    \put(75,  2){\includegraphics*[width=75mm]{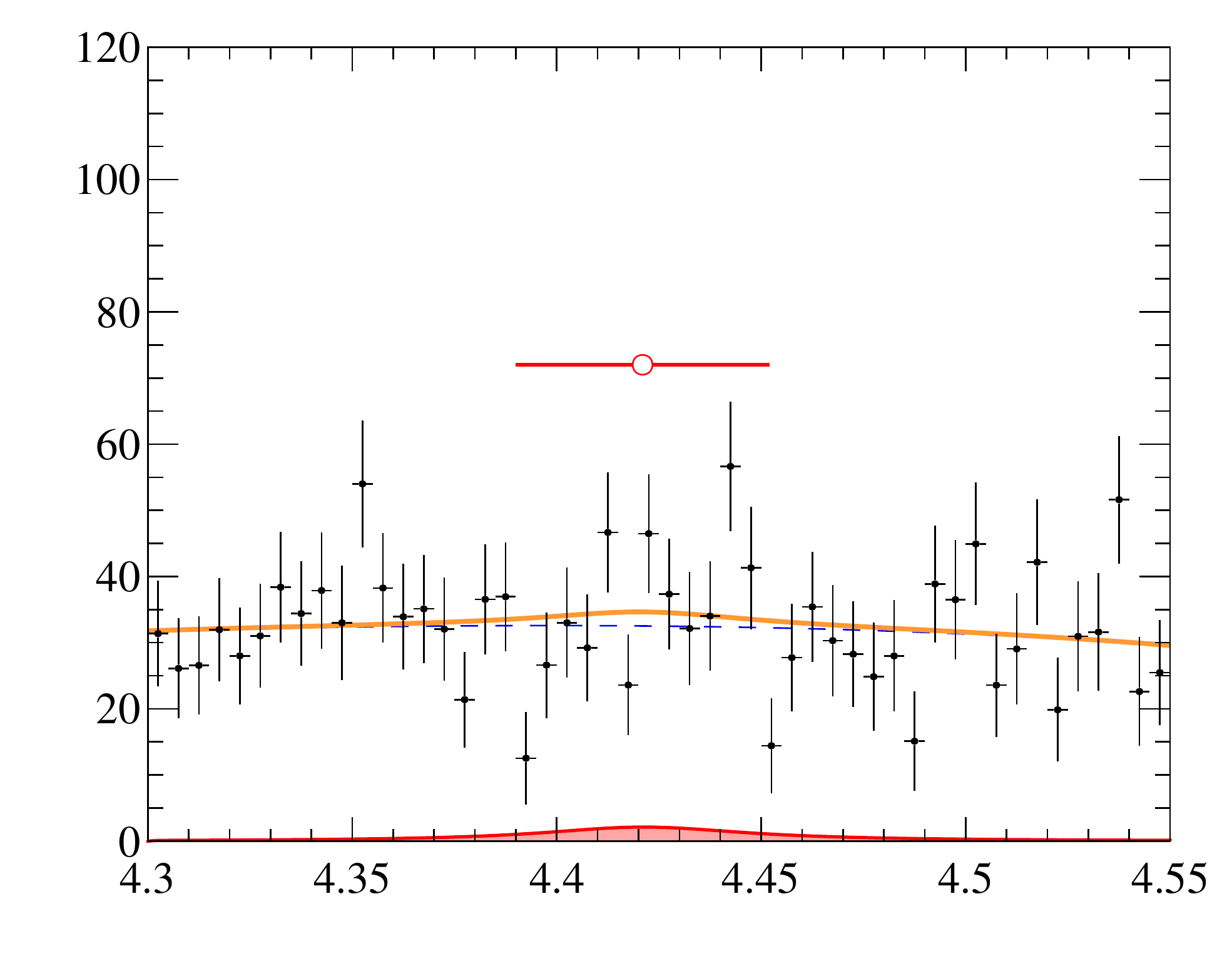}}
    
    \put(  74,151){\begin{sideways}{Yield/$(5\mevcc)$}\end{sideways}}
    \put(  74, 89){\begin{sideways}{Yield/$(12\mevcc)$}\end{sideways}}
    \put(  74, 27){\begin{sideways}{Yield/$(5\mevcc)$}\end{sideways}}
    \put(  -1,151){\begin{sideways}{Yield/$(5\mevcc)$}\end{sideways}}
    \put(  -1, 89){\begin{sideways}{Yield/$(5\mevcc)$}\end{sideways}}
    \put(  -1, 27){\begin{sideways}{Yield/$(5\mevcc)$}\end{sideways}}
    
    \put(35  ,  0){\large$m_{\jpsi \Peta}$}
    \put(35  , 62){\large$m_{\jpsi \Peta}$}
    \put(35  ,124){\large$m_{\jpsi \Peta}$}
    \put(110 ,  0){\large$m_{\jpsi \Peta}$}
    \put(110 , 62){\large$m_{\jpsi \Peta}$}
    \put(110 ,124){\large$m_{\jpsi \Peta}$}

    \put(53 ,  0){\large$\left[\!\gevcc\right]$}
    \put(53 , 62){\large$\left[\!\gevcc\right]$}
    \put(53 ,124){\large$\left[\!\gevcc\right]$}
    \put(128,  0){\large$\left[\!\gevcc\right]$}
    \put(128, 62){\large$\left[\!\gevcc\right]$}
    \put(128,124){\large$\left[\!\gevcc\right]$}
    
    \put( 56,172){$\begin{array}{l}\lhcb \\ 9\invfb\end{array}$}
    \put( 56,110){$\begin{array}{l}\lhcb \\ 9\invfb\end{array}$}
    \put( 56, 48){$\begin{array}{l}\lhcb \\ 9\invfb\end{array}$}
    \put(131,172){$\begin{array}{l}\lhcb \\ 9\invfb\end{array}$}
    \put(131,110){$\begin{array}{l}\lhcb \\ 9\invfb\end{array}$}
    \put(131, 48){$\begin{array}{l}\lhcb \\ 9\invfb\end{array}$}
 
     \put(25,158){\footnotesize$\begin{array}{cl}	 
     \bigplus\mkern-18mu\bullet\ \  & \mathrm{Data} 
     \\
      \begin{tikzpicture}[x=1mm,y=1mm]\filldraw[fill=red!35!white,draw=red,thick]  (0,0) rectangle (8,3);\end{tikzpicture} 
      & \decay{\Bp}{\Ppsi(3770)\Kp}
      \\
     \begin{tikzpicture}[x=1mm,y=1mm]\filldraw[fill=blue!35!white,draw=blue,thick]  (0,0) rectangle (8,3);\end{tikzpicture} 
      & \decay{\Bp}{\psitwos\Kp}
      \\
     {\color{blue}{\hdashrule[0.0ex][x]{8mm}{1.0pt}{2.0mm 0.3mm}}}
      &  \decay{\Bp}{\left(\jpsi\Peta\right)_{\mathrm{NR}}\Kp}
      \\
      {\color[RGB]{255,153,51} {\rule{8mm}{2.0pt}}}
      & \mathrm{Total}
     \end{array}$}
     
     \put(100,158){\footnotesize$\begin{array}{cl}	 
     \bigplus\mkern-18mu\bullet\ \  & \mathrm{Data} 
     \\
      \begin{tikzpicture}[x=1mm,y=1mm]\filldraw[fill=red!35!white,draw=red,thick]  (0,0) rectangle (8,3);\end{tikzpicture} 
      & \decay{\Bp}{\Ppsi_2(3823)\Kp}
      \\
     \begin{tikzpicture}[x=1mm,y=1mm]\filldraw[fill=blue!35!white,draw=blue,thick]  (0,0) rectangle (8,3);\end{tikzpicture} 
      & \decay{\Bp}{\psitwos\Kp}
      \\
     {\color{blue}{\hdashrule[0.0ex][x]{8mm}{1.0pt}{2.0mm 0.3mm}}}
      &  \decay{\Bp}{\left(\jpsi\Peta\right)_{\mathrm{NR}}\Kp}
      \\
      {\color[RGB]{255,153,51} {\rule{8mm}{2.0pt}}}
      & \mathrm{Total}
     \end{array}$}
 
     \put(11,109){\footnotesize$\begin{array}{cl}	 
     \bigplus\mkern-18mu\bullet\ \  & \mathrm{Data} 
     \\
      \begin{tikzpicture}[x=1mm,y=1mm]\filldraw[fill=red!35!white,draw=red,thick]  (0,0) rectangle (8,3);\end{tikzpicture} 
      & \decay{\Bp}{\Ppsi_3(3842)\Kp}
      \\
     {\color{blue}{\hdashrule[0.0ex][x]{8mm}{1.0pt}{2.0mm 0.3mm}}}
      &  \decay{\Bp}{\left(\jpsi\Peta\right)_{\mathrm{NR}}\Kp}
      \\
      {\color[RGB]{255,153,51} {\rule{8mm}{2.0pt}}}
      & \mathrm{Total}
     \end{array}$}

     \put(84,109){\footnotesize$\begin{array}{cl}	 
     \bigplus\mkern-18mu\bullet\ \  & \mathrm{Data} 
     \\
      \begin{tikzpicture}[x=1mm,y=1mm]\filldraw[fill=red!35!white,draw=red,thick]  (0,0) rectangle (8,3);\end{tikzpicture} 
      & \decay{\Bp}{\Ppsi(4040)\Kp}
      \\
     {\color{blue}{\hdashrule[0.0ex][x]{8mm}{1.0pt}{2.0mm 0.3mm}}}
      &  \decay{\Bp}{\left(\jpsi\Peta\right)_{\mathrm{NR}}\Kp}
      \\
      {\color[RGB]{255,153,51} {\rule{8mm}{2.0pt}}}
      & \mathrm{Total}
     \end{array}$}
     \put(12, 46){\footnotesize$\begin{array}{cl}	 
     \bigplus\mkern-18mu\bullet\ \  & \mathrm{Data} 
     \\
      \begin{tikzpicture}[x=1mm,y=1mm]\filldraw[fill=red!35!white,draw=red,thick]  (0,0) rectangle (8,3);\end{tikzpicture} 
      & \decay{\Bp}{\Ppsi(4160)\Kp}
      \\
     {\color{blue}{\hdashrule[0.0ex][x]{8mm}{1.0pt}{2.0mm 0.3mm}}}
      &  \decay{\Bp}{\left(\jpsi\Peta\right)_{\mathrm{NR}}\Kp}
      \\
      {\color[RGB]{255,153,51} {\rule{8mm}{2.0pt}}}
      & \mathrm{Total}
     \end{array}$}
     
     \put(87, 46){\footnotesize$\begin{array}{cl}	 
     \bigplus\mkern-18mu\bullet\ \  & \mathrm{Data} 
     \\
      \begin{tikzpicture}[x=1mm,y=1mm]\filldraw[fill=red!35!white,draw=red,thick]  (0,0) rectangle (8,3);\end{tikzpicture} 
      & \decay{\Bp}{\Ppsi(4415)\Kp}
      \\
     {\color{blue}{\hdashrule[0.0ex][x]{8mm}{1.0pt}{2.0mm 0.3mm}}}
      &  \decay{\Bp}{\left(\jpsi\Peta\right)_{\mathrm{NR}}\Kp}
      \\
      {\color[RGB]{255,153,51} {\rule{8mm}{2.0pt}}}
      & \mathrm{Total}
     \end{array}$}

  \end{picture}
  \caption {\small 
    Background-subtracted 
    $\jpsi\Peta$~mass distribution 
    from~\mbox{$\decay{\Bp}{\jpsi\Peta\Kp}$}~decays 
    in the~vicinity of 
    the~conventional 
    (top row)~$\Ppsi(3770)$, 
    $\Ppsi_2(3823)$, 
    (middle row)~$\Ppsi_3(3842)$, 
    $\Ppsi(4040)$,
    (bottom row)~$\Ppsi(4160)$
     and 
    $\Ppsi(4415)$~charmonium states.
    The~results of the~fits, described 
	in the~text, are overlaid.
    The~red open  point with horizontal error bars indicates 
    the~mass and width of the~resonance 
    assumed in the~fits. 
    }
  \label{fig:signal_fit1}
\end{figure}

\begin{figure}[t]
  \setlength{\unitlength}{1mm}
  \centering
  \begin{picture}(150,185)
    \definecolor{root8}{rgb}{0.35, 0.83, 0.33}
    \put( 0,126){\includegraphics*[width=75mm]{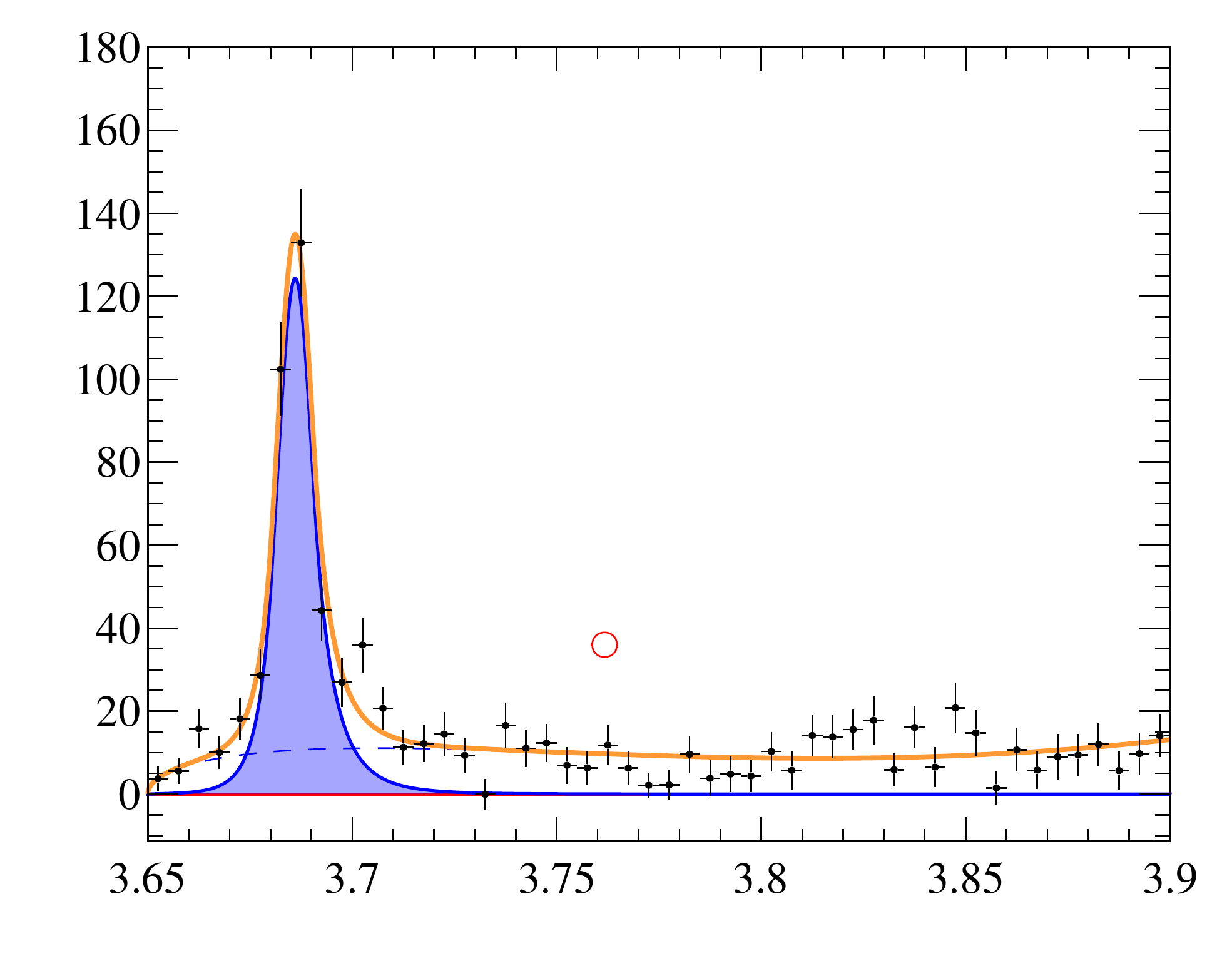}}
    \put(75,126){\includegraphics*[width=75mm]{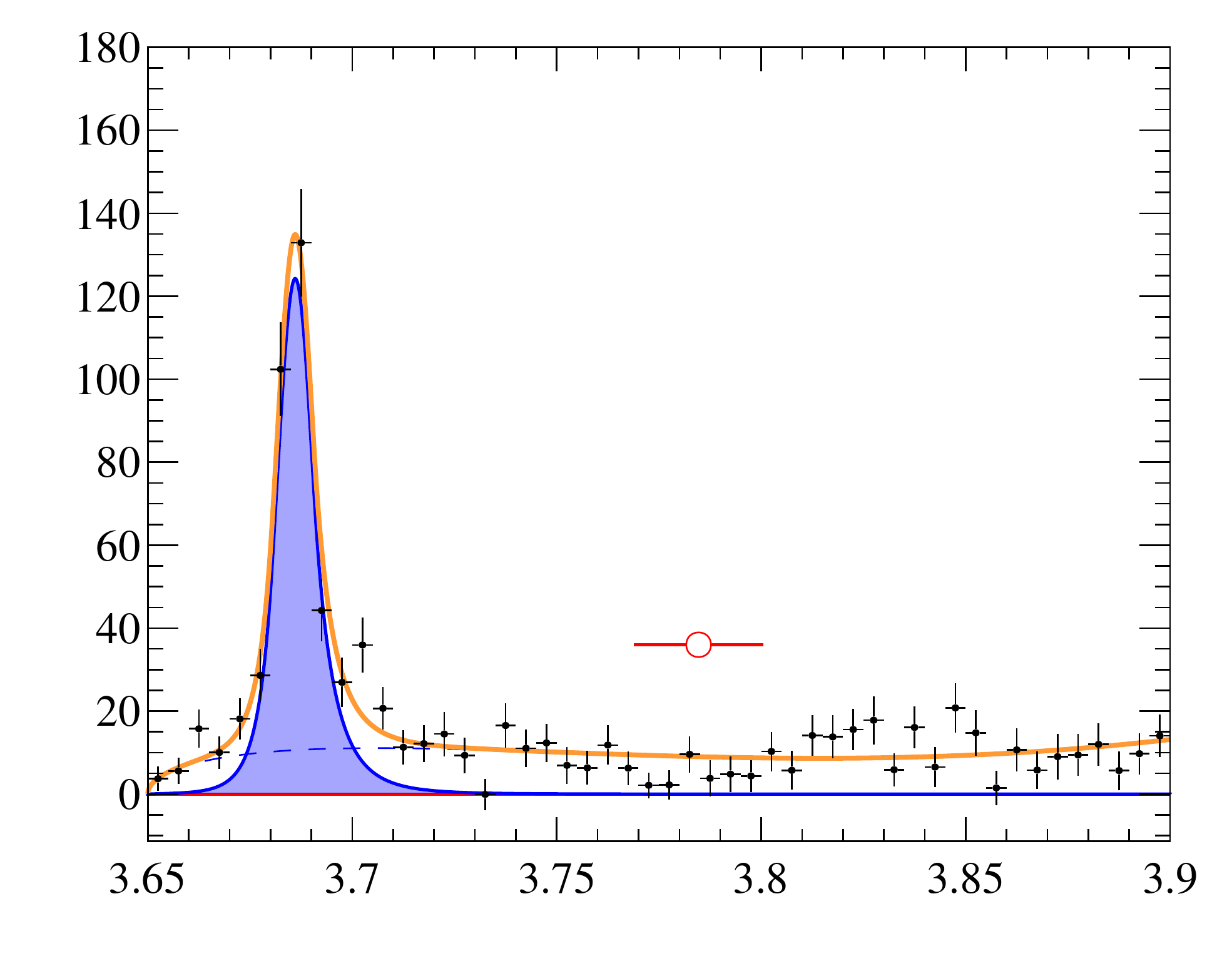}}
    \put( 0, 64){\includegraphics*[width=75mm]{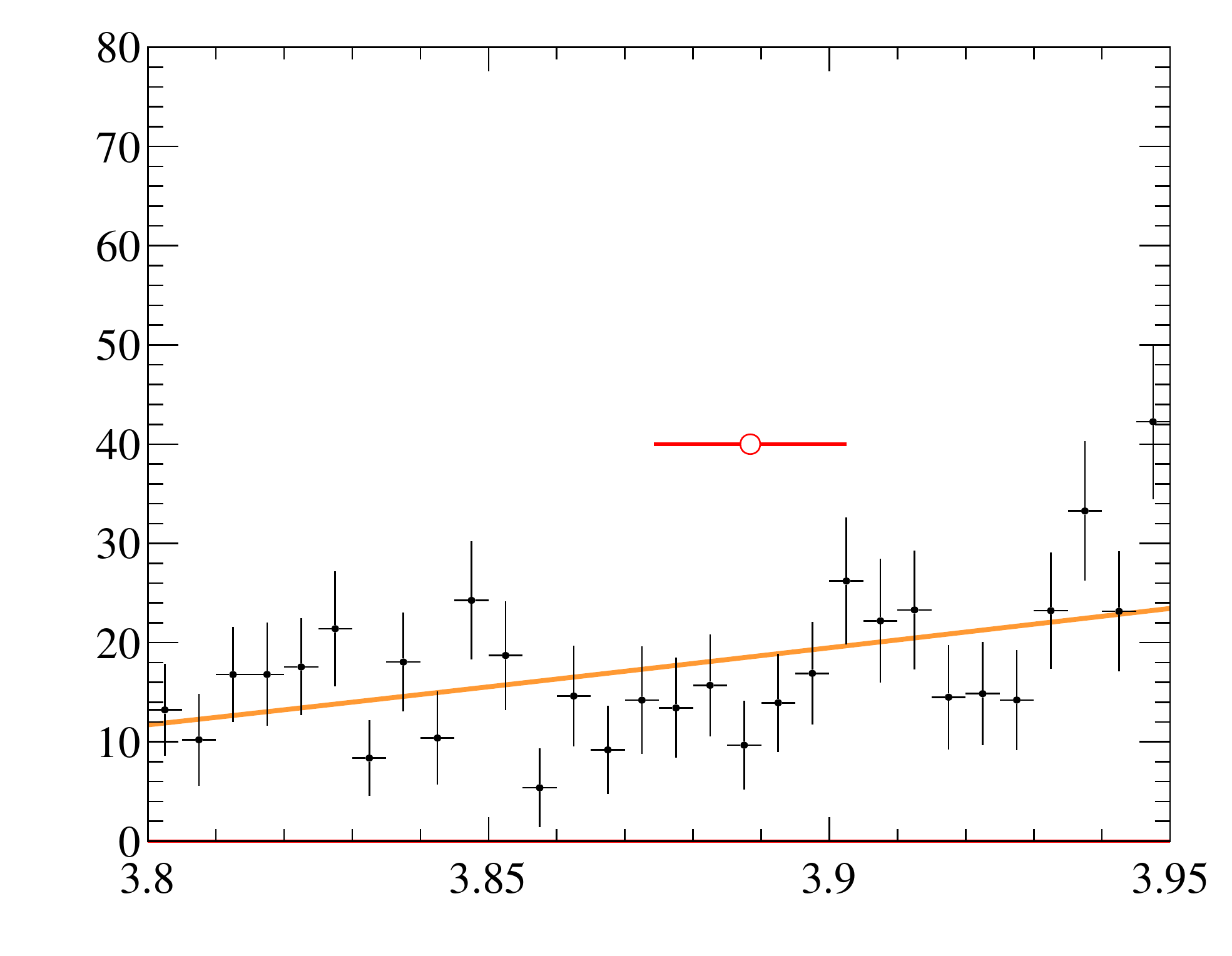}}
    \put(75, 64){\includegraphics*[width=75mm]{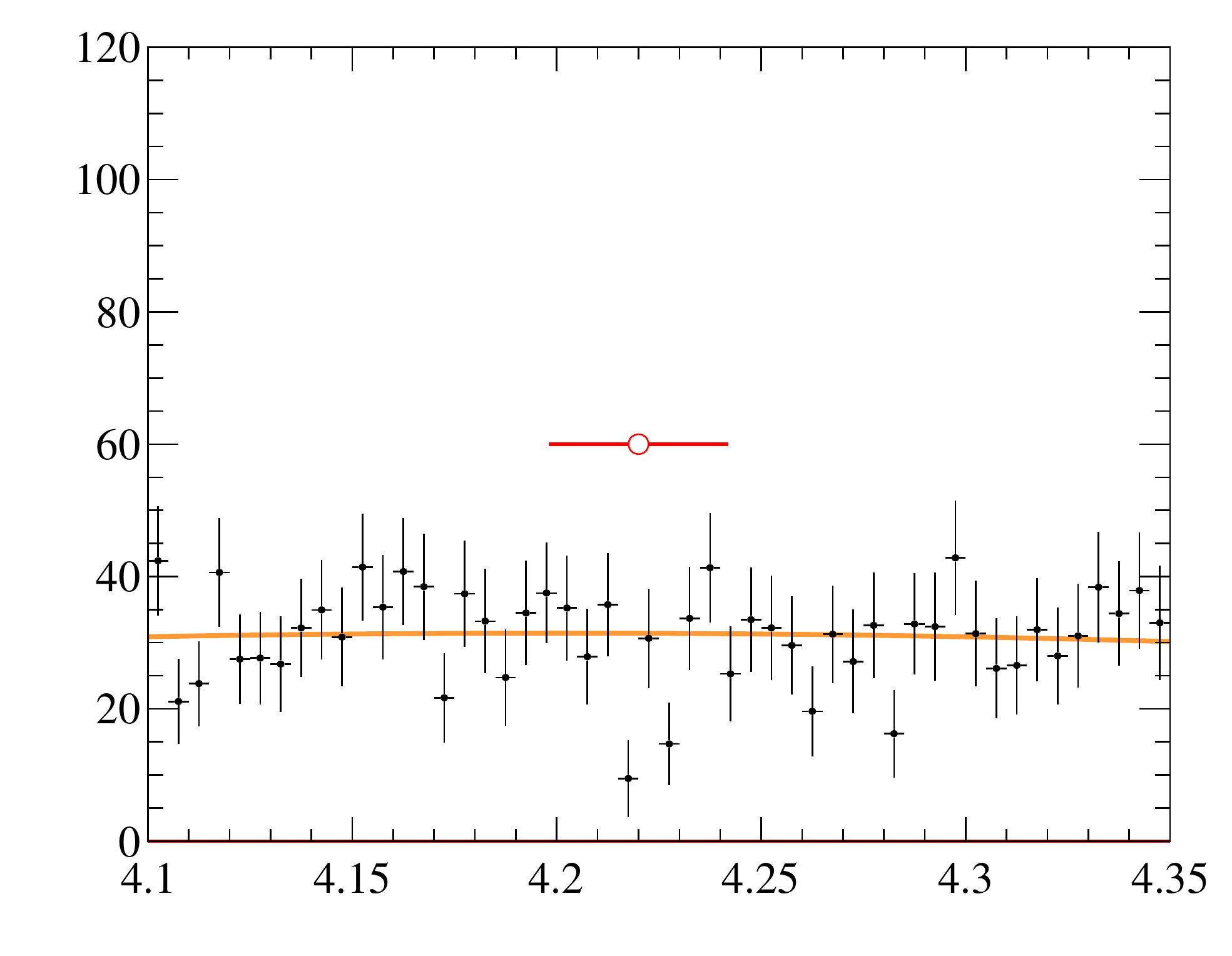}}
    \put( 0,  2){\includegraphics*[width=75mm]{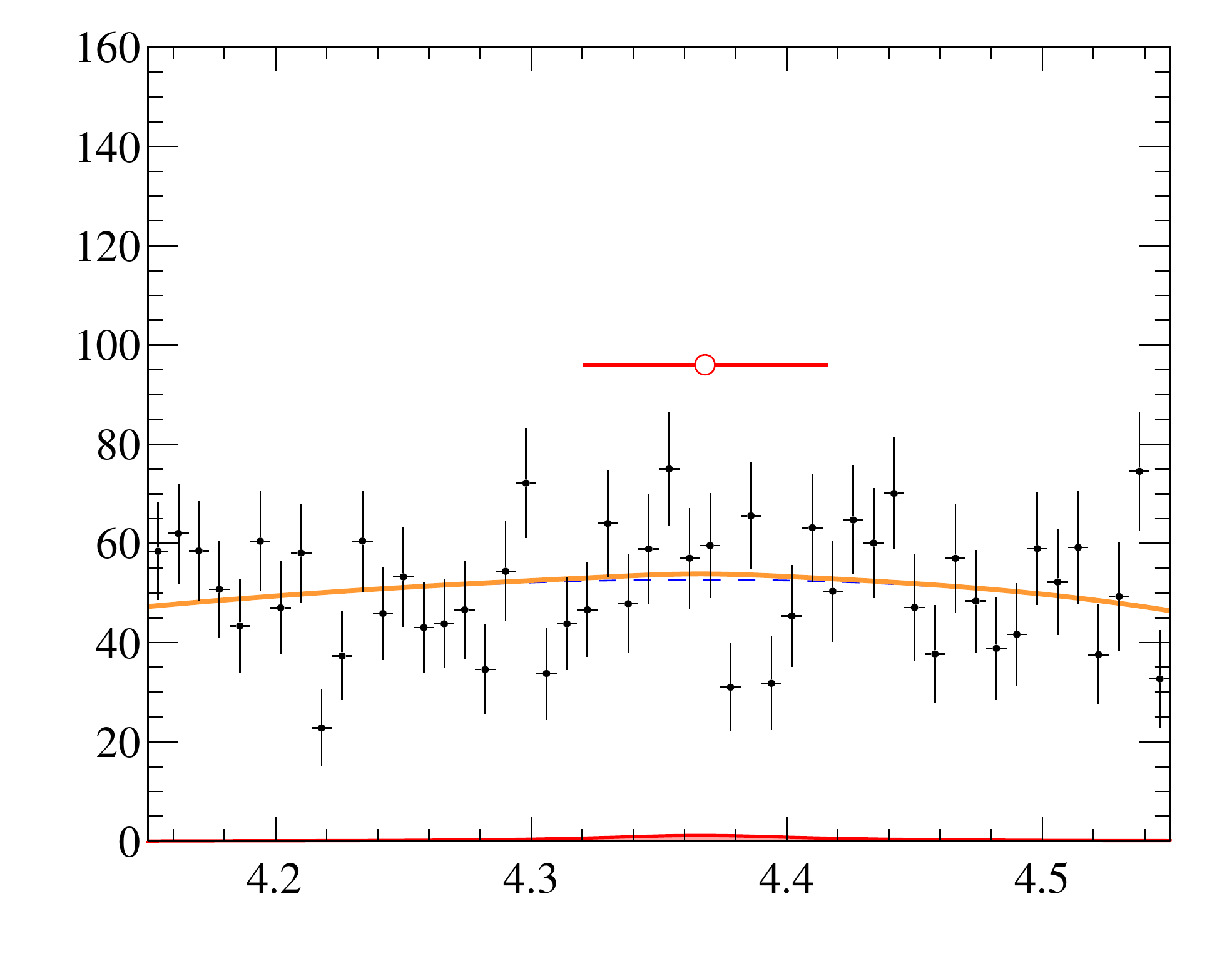}}
    \put(75,  2){\includegraphics*[width=75mm]{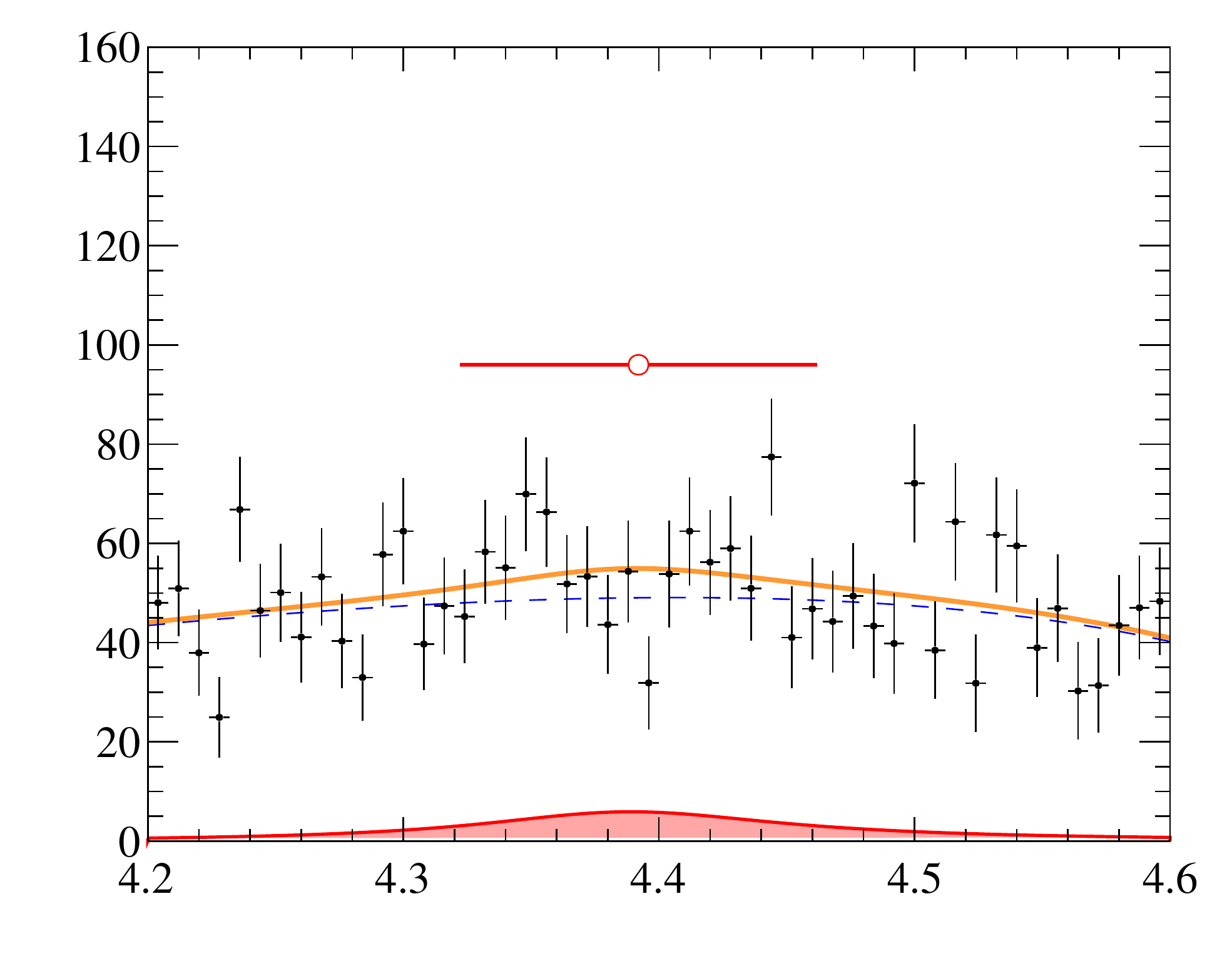}}
    \put(  74,151){\begin{sideways}{Yield/$(5\mevcc)$}\end{sideways}}
    \put(  74, 89){\begin{sideways}{Yield/$(5\mevcc)$}\end{sideways}}
    \put(  74, 27){\begin{sideways}{Yield/$(8\mevcc)$}\end{sideways}}
    \put(  -1,151){\begin{sideways}{Yield/$(5\mevcc)$}\end{sideways}}
    \put(  -1, 89){\begin{sideways}{Yield/$(5\mevcc)$}\end{sideways}}
    \put(  -1, 27){\begin{sideways}{Yield/$(8\mevcc)$}\end{sideways}}
    
    \put(35  ,  0){\large$m_{\jpsi \Peta}$}
    \put(35  , 62){\large$m_{\jpsi \Peta}$}
    \put(35  ,124){\large$m_{\jpsi \Peta}$}
    \put(110 ,  0){\large$m_{\jpsi \Peta}$}
    \put(110 , 62){\large$m_{\jpsi \Peta}$}
    \put(110 ,124){\large$m_{\jpsi \Peta}$}

    \put(53 ,  0){\large$\left[\!\gevcc\right]$}
    \put(53 , 62){\large$\left[\!\gevcc\right]$}
    \put(53 ,124){\large$\left[\!\gevcc\right]$}
    \put(128,  0){\large$\left[\!\gevcc\right]$}
    \put(128, 62){\large$\left[\!\gevcc\right]$}
    \put(128,124){\large$\left[\!\gevcc\right]$}
    
    \put( 56,172){$\begin{array}{l}\lhcb \\ 9\invfb\end{array}$}
    \put( 56,110){$\begin{array}{l}\lhcb \\ 9\invfb\end{array}$}
    \put( 56, 48){$\begin{array}{l}\lhcb \\ 9\invfb\end{array}$}
    \put(131,172){$\begin{array}{l}\lhcb \\ 9\invfb\end{array}$}
    \put(131,110){$\begin{array}{l}\lhcb \\ 9\invfb\end{array}$}
    \put(131, 48){$\begin{array}{l}\lhcb \\ 9\invfb\end{array}$}
 
     \put(25,158){\footnotesize$\begin{array}{cl}	 
     \bigplus\mkern-18mu\bullet\ \  & \mathrm{Data} 
     \\
      \begin{tikzpicture}[x=1mm,y=1mm]\filldraw[fill=red!35!white,draw=red,thick]  (0,0) rectangle (8,3);\end{tikzpicture} 
      & \decay{\Bp}{\PR(3760)\Kp}
      \\
      \begin{tikzpicture}[x=1mm,y=1mm]\filldraw[fill=blue!35!white,draw=blue,thick]  (0,0) rectangle (8,3);\end{tikzpicture}
      & \decay{\Bp}{\psitwos\Kp}
      \\
     {\color{blue}{\hdashrule[0.0ex][x]{8mm}{1.0pt}{2.0mm 0.3mm}}}
      &  \decay{\Bp}{\left(\jpsi\Peta\right)_{\mathrm{NR}}\Kp}
      \\
      {\color[RGB]{255,153,51} {\rule{8mm}{2.0pt}}}
      & \mathrm{Total}
     \end{array}$}
     
     \put(102,158){\footnotesize$\begin{array}{cl}	 
     \bigplus\mkern-18mu\bullet\ \  & \mathrm{Data} 
     \\
      \begin{tikzpicture}[x=1mm,y=1mm]\filldraw[fill=red!35!white,draw=red,thick]  (0,0) rectangle (8,3);\end{tikzpicture} 
      & \decay{\Bp}{\PR(3790)\Kp}
      \\
      \begin{tikzpicture}[x=1mm,y=1mm]\filldraw[fill=blue!35!white,draw=blue,thick]  (0,0) rectangle (8,3);\end{tikzpicture} 
      & \decay{\Bp}{\psitwos\Kp}
      \\
     {\color{blue}{\hdashrule[0.0ex][x]{8mm}{1.0pt}{2.0mm 0.3mm}}}
      &  \decay{\Bp}{\left(\jpsi\Peta\right)_{\mathrm{NR}}\Kp}
      \\
      {\color[RGB]{255,153,51} {\rule{8mm}{2.0pt}}}
      & \mathrm{Total}
     \end{array}$}
 
     \put(11,108){\footnotesize$\begin{array}{cl}	 
     \bigplus\mkern-18mu\bullet\ \  & \mathrm{Data} 
     \\
      \begin{tikzpicture}[x=1mm,y=1mm]\filldraw[fill=red!35!white,draw=red,thick]  (0,0) rectangle (8,3);\end{tikzpicture} 
      & \decay{\Bp}{\Z_{\cquark}(3900)^0\Kp}
      \\
     {\color{blue}{\hdashrule[0.0ex][x]{8mm}{1.0pt}{2.0mm 0.3mm}}}
      &  \decay{\Bp}{\left(\jpsi\Peta\right)_{\mathrm{NR}}\Kp}
      \\
      {\color[RGB]{255,153,51} {\rule{8mm}{2.0pt}}}
      & \mathrm{Total}
     \end{array}$}

     \put(86,108){\footnotesize$\begin{array}{cl}	 
     \bigplus\mkern-18mu\bullet\ \  & \mathrm{Data} 
     \\
      \begin{tikzpicture}[x=1mm,y=1mm]\filldraw[fill=red!35!white,draw=red,thick]  (0,0) rectangle (8,3);\end{tikzpicture} 
      & \decay{\Bp}{\Ppsi(4230)\Kp}
      \\
     {\color{blue}{\hdashrule[0.0ex][x]{8mm}{1.0pt}{2.0mm 0.3mm}}}
      &  \decay{\Bp}{\left(\jpsi\Peta\right)_{\mathrm{NR}}\Kp}
      \\
      {\color[RGB]{255,153,51} {\rule{8mm}{2.0pt}}}
      & \mathrm{Total}
     \end{array}$}

     \put(11, 46){\footnotesize$\begin{array}{cl}	 
     \bigplus\mkern-18mu\bullet\ \  & \mathrm{Data} 
     \\
      \begin{tikzpicture}[x=1mm,y=1mm]\filldraw[fill=red!35!white,draw=red,thick]  (0,0) rectangle (8,3);\end{tikzpicture} 
      & \decay{\Bp}{\Ppsi(4360)\Kp}
      \\
     {\color{blue}{\hdashrule[0.0ex][x]{8mm}{1.0pt}{2.0mm 0.3mm}}}
      &  \decay{\Bp}{\left(\jpsi\Peta\right)_{\mathrm{NR}}\Kp}
      \\
      {\color[RGB]{255,153,51} {\rule{8mm}{2.0pt}}}
      & \mathrm{Total}
     \end{array}$}
     
     \put(86, 46){\footnotesize$\begin{array}{cl}	 
     \bigplus\mkern-18mu\bullet\ \  & \mathrm{Data} 
     \\
      \begin{tikzpicture}[x=1mm,y=1mm]\filldraw[fill=red!35!white,draw=red,thick]  (0,0) rectangle (8,3);\end{tikzpicture} 
      & \decay{\Bp}{\Ppsi(4390)\Kp}
      \\
     {\color{blue}{\hdashrule[0.0ex][x]{8mm}{1.0pt}{2.0mm 0.3mm}}}
      &  \decay{\Bp}{\left(\jpsi\Peta\right)_{\mathrm{NR}}\Kp}
      \\
      {\color[RGB]{255,153,51} {\rule{8mm}{2.0pt}}}
      & \mathrm{Total}
     \end{array}$}

  \end{picture}
  \caption {\small 
    Background-subtracted 
    $\jpsi\Peta$~mass distribution 
    from~\mbox{$\decay{\Bp}{\jpsi\Peta\Kp}$}~decays 
    in the~vicinity of 
    the~charmonum\protect\nobreakdash-like 
    (top row)~$\PR(3760)$,
    $\PR(3790)$, 
    (middle row)~$\PZ(3900)^0$,
    $\Ppsi(4230)$, 
    (bottom row)~$\Ppsi(4360)$ and 
    $\Ppsi(4390)$~states.
    The~results of the~fits, described 
	in the~text, are overlaid.
    The~red open  point with horizontal error bars indicates 
    the~mass and width of the~resonance 
    assumed in the~fits. 
    }
  \label{fig:signal_fit2}
\end{figure}

\begin{figure}[t]
  \setlength{\unitlength}{1mm}
  \centering
   \begin{picture}(150,61)
    \definecolor{root8}{rgb}{0.35, 0.83, 0.33}
    \put( 0, 2){\includegraphics*[width=75mm]{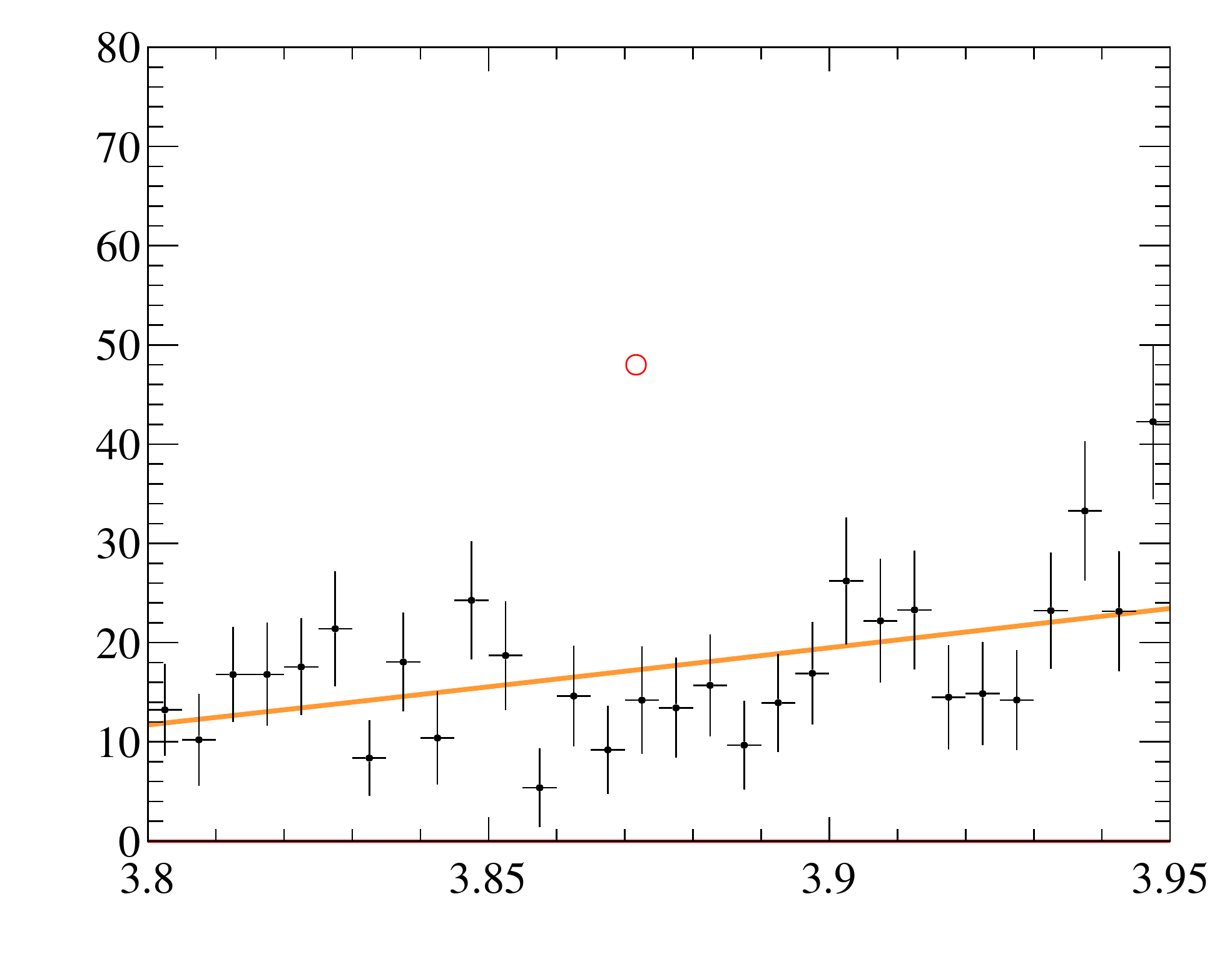}}
    \put(75, 2){\includegraphics*[width=75mm]{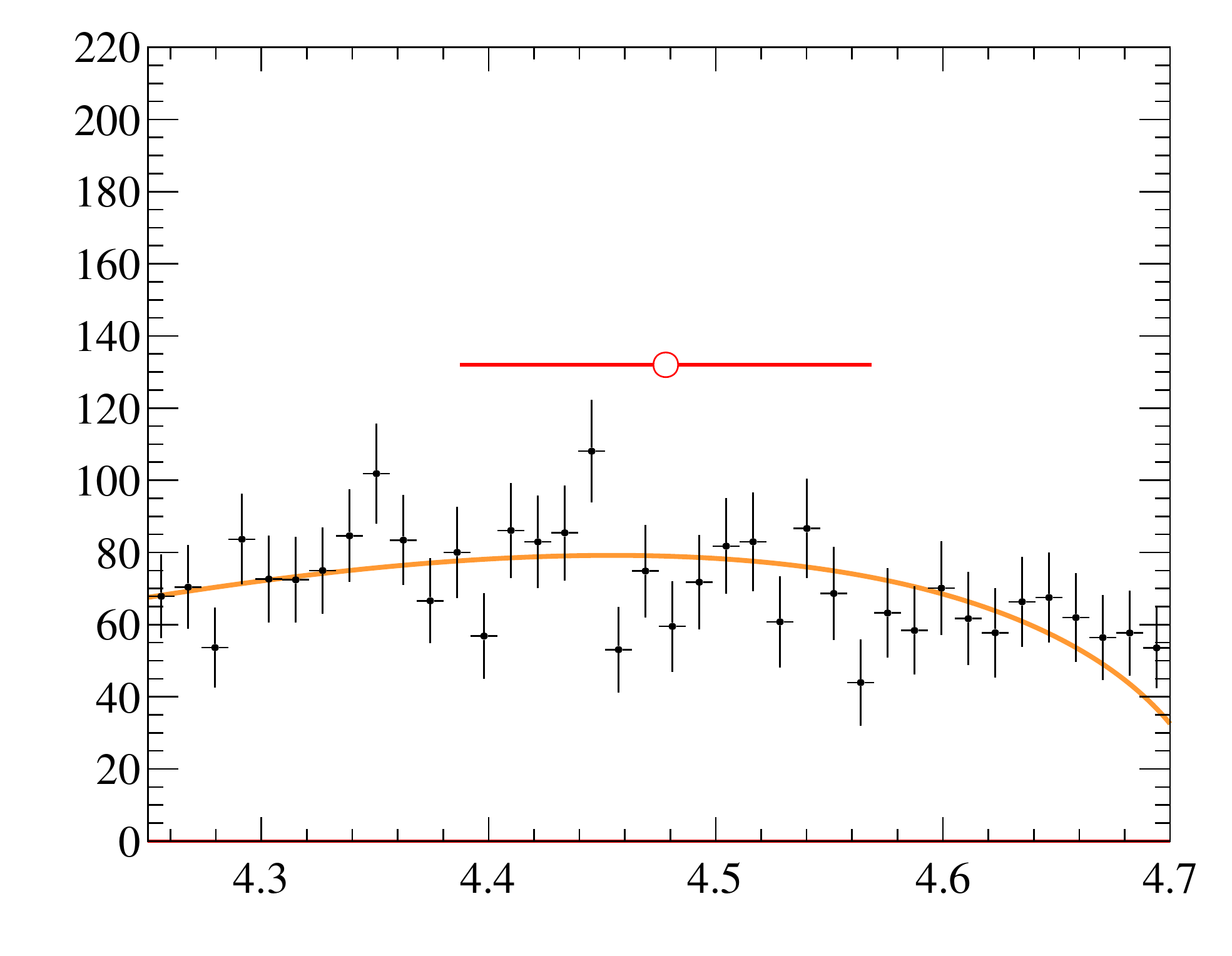}}
    \put(  -1, 27){\begin{sideways}{Yield/$(5\mevcc)$}\end{sideways}}
    \put(  74, 27){\begin{sideways}{Yield/$(12\mevcc)$}\end{sideways}}
    
    \put(35  ,  0){\large$m_{\jpsi \Peta}$}
    \put(110 ,  0){\large$m_{\jpsi \Peta}$}

    \put(53 ,  0){\large$\left[\!\gevcc\right]$}
    \put(128,  0){\large$\left[\!\gevcc\right]$}

    \put( 56, 48){$\begin{array}{l}\lhcb \\ 9\invfb\end{array}$}
    \put(131, 48){$\begin{array}{l}\lhcb \\ 9\invfb\end{array}$}

     \put(11,46){\footnotesize$\begin{array}{cl}	 
     \bigplus\mkern-18mu\bullet\ \  & \mathrm{Data} 
     \\
      \begin{tikzpicture}[x=1mm,y=1mm]\filldraw[fill=red!35!white,draw=red,thick]  (0,0) rectangle (8,3);\end{tikzpicture} 
      & \decay{\Bp}{\PX^{\prime}_{\PC}\Kp}
      \\
     {\color{blue}{\hdashrule[0.0ex][x]{8mm}{1.0pt}{2.0mm 0.3mm}}}
      &  \decay{\Bp}{\left(\jpsi\Peta\right)_{\mathrm{NR}}\Kp}
      \\
      {\color[RGB]{255,153,51} {\rule{8mm}{2.0pt}}}
      & \mathrm{Total}
     \end{array}$}

     \put(86,46){\footnotesize$\begin{array}{cl}	 
     \bigplus\mkern-18mu\bullet\ \  & \mathrm{Data} 
     \\
      \begin{tikzpicture}[x=1mm,y=1mm]\filldraw[fill=red!35!white,draw=red,thick]  (0,0) rectangle (8,3);\end{tikzpicture} 
      & \decay{\Bp}{\Z_{\cquark}(4430)^0\Kp}
      \\
     {\color{blue}{\hdashrule[0.0ex][x]{8mm}{1.0pt}{2.0mm 0.3mm}}}
      &  \decay{\Bp}{\left(\jpsi\Peta\right)_{\mathrm{NR}}\Kp}
      \\
      {\color[RGB]{255,153,51} {\rule{8mm}{2.0pt}}}
      & \mathrm{Total}
     \end{array}$}
  \end{picture}
  \caption {\small 
      Background-subtracted 
    $\jpsi\Peta$~mass distribution 
    from~\mbox{$\decay{\Bp}{\jpsi\Peta\Kp}$}~decays 
    in the~vicinity of 
    the~hypothetical 
    (left)~$\PX^{\prime}_{\PC}$
    and (right)~$\Z_{\cquark}(4430)^0$~states. 
     The~results of the~fits, described 
	in the~text, are overlaid.
    The~red open  point with horizontal error bars indicates 
    the~mass and width of the~resonance used in the~fits. 
   }
  \label{fig:signal_fit3}
\end{figure}
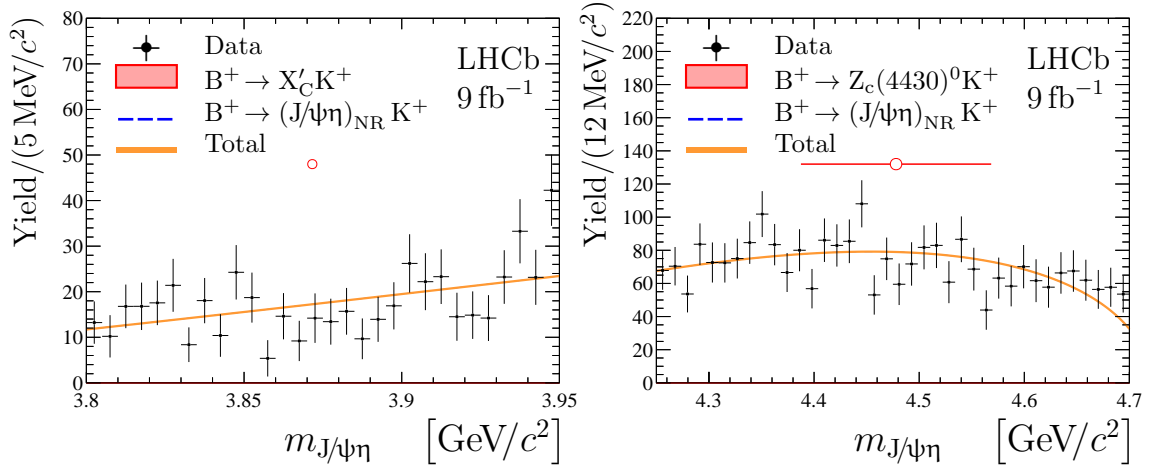

In addition to the~search for  
decays
with a~narrow hypothetical \PX state, 
a~search is performed for decays mediated by
known conventional 
charmonium   or 
charmonium\nobreakdash-like states, 
including the~hypothetical $\PX^{\prime}_{\PC}$~state 
and the neutral partner of 
the~$\PZ_{\cquark}(4430)^+$~state. 
For~the~latter 
it is assumed that the~mass and width 
are the~same as for its 
charged partner~\cite{Belle:2007hrb,
Belle:2013shl,
LHCb-PAPER-2014-014,
Belle:2014nuw,
LHCb-PAPER-2015-038},
 while for the~$\PX^{\prime}_{\PC}$~state 
 the~mass and width are assumed to be 
 the~same as for
 the~$\chicone(3872)$~state~\cite{LHCb-PAPER-2020-008,
 LHCb-PAPER-2020-009}.
The~$\mathcal{C}_{\PX}$~component 
is parametrised 
with a~relativistic $\PS$\nobreakdash-wave Breit\nobreakdash--Wigner
shape convolved with 
the~$\mathcal{F}_S$~function.
 For~each resonance with a~mass
 larger than 3.9\gevcc
 a~fit range is chosen individually depending on 
 the~resonance mass and width.
 The~uncertainties on the~resonance parameters
 are included in the~fits using Gaussian constraints. 
For~the~$\Ppsi_2(3823)$~state, where 
only the~upper limit for the~natural width 
is known~\cite{LHCb-PAPER-2020-009},
a~natural width of 1\mev is assumed. 
The~background\nobreakdash-subtracted
$\jpsi\Peta$~mass spectra in 
the~corresponding ranges, 
together 
with the~fit results, 
are shown in 
Figs.~\ref{fig:signal_fit1} to~\ref{fig:signal_fit3}.
For the~\mbox{$\decay{\Bu}
{\left(\decay{\Ppsi_2(3823)}{\jpsi\Peta}\right)\Kp}$}
and \mbox{$\decay{\Bu}{\left(
\decay{\Ppsi(4040)}{\jpsi\Peta}\right) \Kp}$}~decays,
signals with a~statistical significance 
of 
3.4 
and 9.0
standard deviations, respectively, 
are seen. 
No~evidence for other decays is obtained.

\section{Efficiency and systematic uncertainty}\label{sec:eff} 

For each considered value of $m_\PX$, 
the~efficiency ratio 
$R_{\varepsilon}$ from Eq.~\eqref{eq:NXa} 
is calculated as 
\begin{equation}
   R_{\varepsilon}(m_{\PX})\equiv 
   \dfrac
   {\varepsilon ({\decay{\Bp}{\left(\PX\to\jpsi\Peta\right)\Kp}})}
   {\varepsilon({\decay{\Bp}{\left(\psitwos\to\jpsi\Peta\right)\Kp}})} \,,
\end{equation}
where the~total efficiency $\varepsilon$ for each decay
is calculated
from the~product of the~detector acceptance,
the~reconstruction and selection efficiencies 
for decays within the~detector acceptance,
and the~trigger efficiency for decays passing 
the~selection criteria.
All~efficiencies are calculated 
using simulation, as described in Sec.~\ref{sec:Detector}.
The~finite size of the~simulation samples 
contributes to the~uncertainty on 
the~$R_{\varepsilon}(m_{\PX})$ ratio.
Since signal and normalisation decays 
share the~same final state, many systematic uncertainties 
cancel in the~ratio $R_{\varepsilon}$.  
The~remaining nonnegligible uncertainties 
are listed in Table~\ref{tab:systematics}.

\begin{table}[b]
	\centering
	\caption{
		 Relative systematic uncertainties for  
		 the~efficiency ratio $R_{\varepsilon}$.
		 When an uncertainty is found to be dependent on the $\jpsi\Peta$ mass, 
         the corresponding range is shown.
		 The~total uncertainty is obtained 
		 as the quadratic sum of the individual contributions.
	}
	\vspace*{3mm}
	\begin{tabular}{lc}
		Source & Uncertainty~$\left[\%\right]$
        \\[2.0mm]
        \hline 
        \\[-2mm]
    Simulation sample size               &   $ 1.0 - 4.0$   \\ 
   \Bp~meson kinematics                  &   $  0.1 $       \\
    Kaon identification                  &   $  1.0 $       \\
    Tracking efficiency correction       &   $  0.02 - 0.15$  \\
    Trigger                              &   $ 1.1 $        \\
    Data-simulation agreement            &   $ 4.0 $    
        \\[2.0mm]
        \hline
        \\[-2mm]
		Sum in quadrature             &  $4.7 - 6.6$
	\end{tabular}
	\vspace*{3mm}
	\label{tab:systematics}
\end{table}
A~large class of systematic uncertainties 
is associated to 
the~corrections applied to the~simulation. 
The~finite size of
the~$\decay{\Bp}{\jpsi\Kp}$ signal sample
used for correction of the~simulated, 
\pt~and $y$~spectra of \Bp~mesons,
induces a~corresponding uncertainty.
In~turn, 
the~variation 
within their uncertainties
induces small changes in 
the~ratio $R_{\varepsilon}$.
The~corresponding spread of these changes amounts to
$0.1\%$ and is taken as systematic uncertainty.

The~kaon identification  variable  used 
for the~{\sc{MLP}}~estimator is  drawn from 
calibration data samples 
accounting for the~dependence 
on particle kinematics and track multiplicity.  
Systematic uncertainties in this procedure arise 
from the~limited statistics of both 
the~simulation and calibration samples, 
and the~modelling of the~identification variable. 
The~limitations due to both simulation and 
calibration sample size are evaluated 
by bootstrapping to create multiple samples, 
and repeating the~procedure for each sample. 
The~impact of potential mismodelling 
of the~kaon identification  variable
is evaluated by describing 
the~corresponding distributions 
using density estimates with 
different kernel widths~\cite{LHCb-DP-2018-001}. 
For~each of these cases, 
alternative efficiency maps are 
produced to determine the~associated uncertainties.
A~systematic uncertainty of 1\% is 
assigned from the~observed differences 
with alternative efficiency maps.

There are residual differences in 
the~reconstruction efficiency 
of charged\nobreakdash-particle tracks that 
do not cancel completely in the~ratio 
$R_{\varepsilon}$
due 
to the slightly~different kinematic distributions 
of the~final\nobreakdash-state particles.
The~track\nobreakdash-finding efficiencies 
obtained from 
simulated samples 
are corrected using
calibration channels~\cite{LHCb-DP-2013-002}.
The~uncertainties related to~the~efficiency 
correction factors 
are propagated to the~ratios of 
the~total efficiencies using pseudoexperiments
and are found to be less than 0.15\%  
for the~considered values of 
the~$m_{\PX}$~parameter.
Differences between data and simulation 
of~photon reconstruction efficiencies
are studied using a~large sample of 
\mbox{$\decay{\Bu}{\jpsi
\left(\decay{\Kstarp}
{\Kp \left(\decay{\piz}{\g\g}\right)}
\right)}$}~decays~\mbox{\cite{LHCb-PAPER-2012-022,
LHCb-PAPER-2012-053,
Govorkova:2015vqa}.}
The~associated systematic 
uncertainty largely cancels.

A~systematic uncertainty  related 
to the~knowledge of the~trigger efficiencies 
has been previously studied using large 
samples of \mbox{$\decay{\Bu}{\left(\decay{\jpsi}{\mumu}\right)\Kp}$} 
and \mbox{$\decay{\Bu}{\left(\decay{\psitwos}{\mumu}\right)\Kp}$}~decays
by comparing the 
ratios of the trigger efficiencies in data 
and simulation~\cite{LHCb-PAPER-2012-010}.
Based on this comparison, a~relative uncertainty 
of 1.1\% is assigned. 

Another possible source of uncertainty
is the~potential disagreement
between data and simulation in the~estimation of efficiencies 
due to effects not considered above. 
This is studied by varying 
the~selection criteria in ranges that lead to
changes in the~measured signal yields 
as large as $\pm 20\%$. 
For~this study,
the~\mbox{$\decay{\Bu}{\left(\decay{\psitwos}
{\jpsi\Peta}\right)\Kp}$}
data sample is used. 
The~resulting difference 
in data\nobreakdash-simulation efficiency ratio 
does not exceed $4.0\%$,
which is conservatively 
taken as systematic uncertainty. 

The~systematic uncertainties discussed 
above affect the ratio of the total efficiencies 
$R_{\varepsilon}$, and are accounted for 
in the~fits using Gaussian constraints.
A~different class of systematic uncertainties 
directly affects the~fit itself, namely 
uncertainties associated with the~fit 
models used
to describe the~$\jpsi\Peta$ and 
$\jpsi\Peta\Kp$~mass spectra.
The~systematic uncertainty is accounted for 
by using fits with alternative models. 
The~alternative resolution models
for the~$\mathcal{C}_{\psitwos}$
and $\mathcal{C}_{\PX}$~components include 
a~generalised 
Student's 
$t$\nobreakdash-distribution~\cite{Student} 
and 
a~sum of two modified Gaussian functions
with a~power\nobreakdash-law tail on 
each side of the~distribution. 
For~wide
charmonia and charmonium\nobreakdash-like 
resonances
a~P\nobreakdash-wave relativistic Breit--Wigner 
function is also tested instead of 
the~S\nobreakdash-wave profile. 
The~tail parameters 
of the~$\mathcal{F}_{\mathcal{S}}$~resolution 
functions are varied within their uncertainties, 
as determined from the~simulation. 
For~the~$C_{\mathrm{NR}}$~component,
the~degree of the~polynomial function is varied between 
zero and two.
For the~signal component of 
the~fit to the~$\jpsi\Peta\Kp$~mass spectrum, 
the~list of alternative models consists of 
a~bifurcated generalised 
Student's $t$-distribution~\cite{Student},
an~Apollonius function~\cite{Santos:2013gra}, 
and  a~sum of two modified Gaussian functions
with a~power\nobreakdash-law tail on 
each side of the~distribution. 
For~the~background component,
the~second\nobreakdash-degree 
positive\nobreakdash-definite polynomial function,
and the~product of an~exponential function 
and a~second\nobreakdash-degree positive\nobreakdash-definite
polynomial function, are tested as 
alternative models. 
For each alternative model a~fit 
to the~$\jpsi\Peta$~mass spectrum is performed
and the~upper limit\,(UL) on the~$F_{\PX}$
or $B_{\PX}$~value is determined 
and conservatively the~largest value of 
the~upper limit is taken to account for 
the~systematic uncertainty. 
For~the~\mbox{$\decay{\Bu}
{\left(\decay{\Ppsi_2(3823)}{\jpsi\Peta}\right)\Kp}$}
and \mbox{$\decay{\Bu}{\left(
\decay{\Ppsi(4040)}{\jpsi\Peta}\right) \Kp}$}~signals,
the~maximal deviation relative to the~baseline fit 
is taken as uncertainty and added in quadrature 
to the~uncertainty obtained from the~fit.
For~the~$\Ppsi_2(3823)$~state, only 
90\,(95)\%\,confidence level\,(CL) upper 
limits on the~natural 
width of 5.2\,(6.6)\mev~are
known~\cite{LHCb-PAPER-2020-009}.
For~this case fits with the~natural 
width varied between 0.2 and 6.6\mev  
are performed  and the~maximal deviation
with respect to the~default fit, 
where 
1\mev
is assumed, 
is taken as the~corresponding 
systematic uncertainty.

\section{Results and summary}

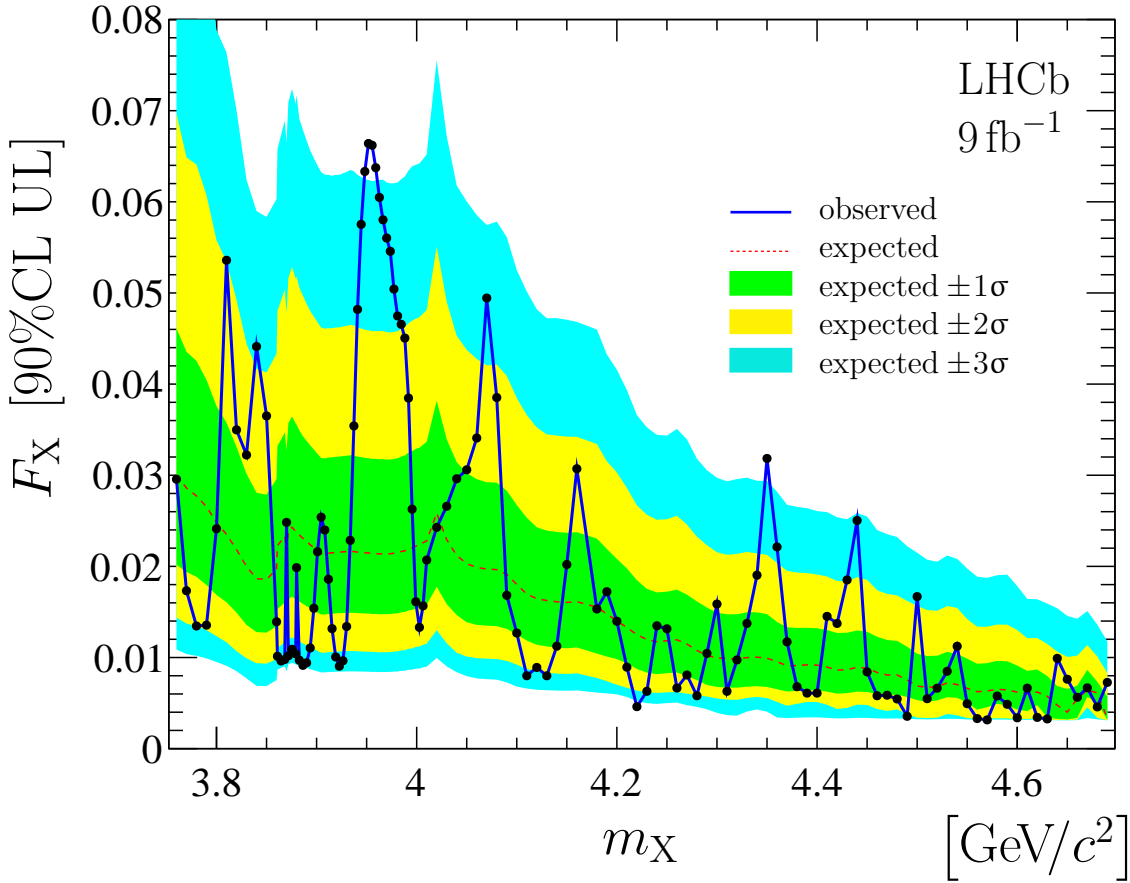
\begin{figure}[t]
	\setlength{\unitlength}{1mm}
	\centering
	\begin{picture}(170,120)
	\put(0,1){\includegraphics*[width=150mm]{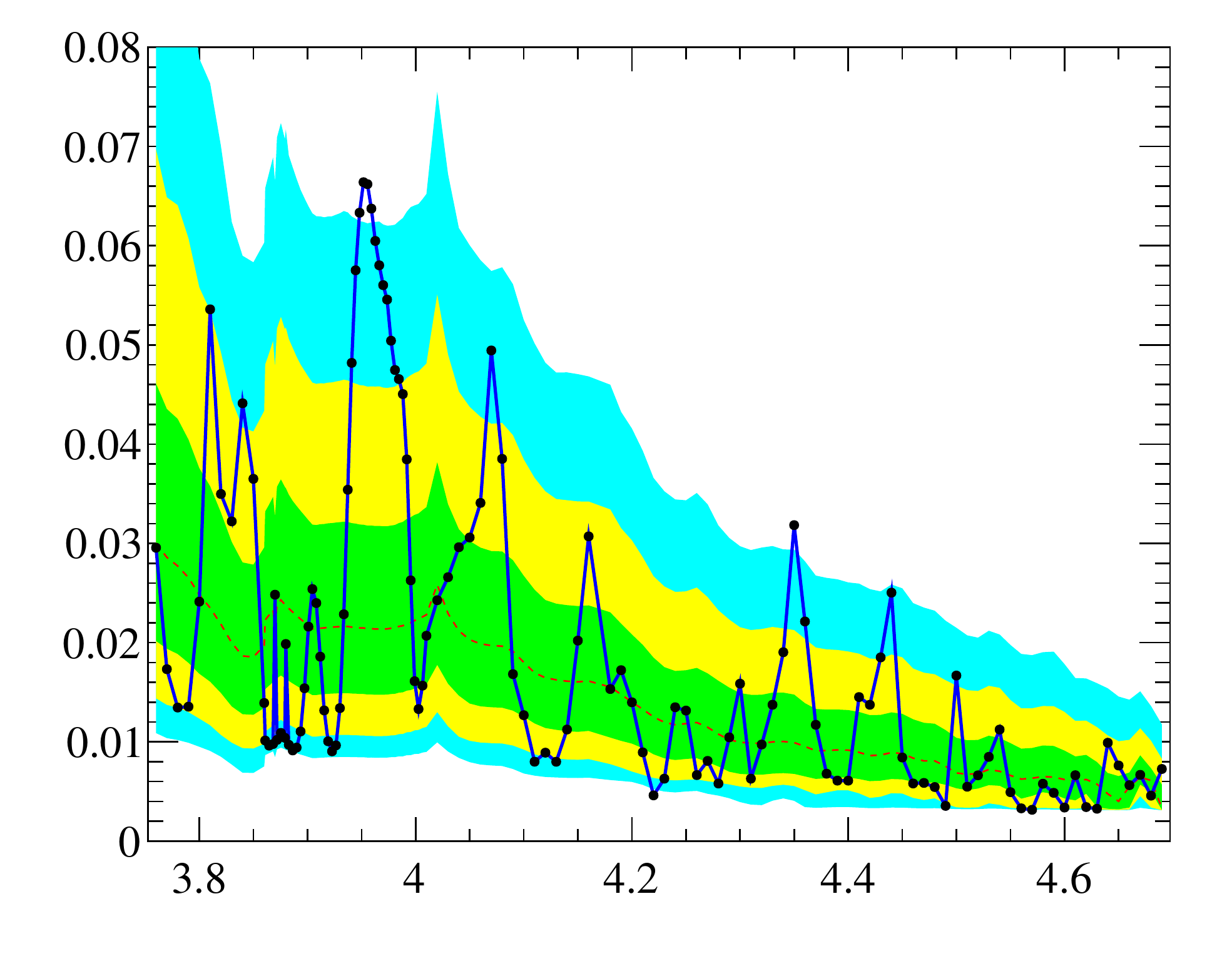}}
        \definecolor{aqua}{rgb}{0.0, 1.0, 1.0}
        \definecolor{brightturquoise}{rgb}{0.03, 0.91, 0.87}
	\put(100,55){\begin{tikzpicture}[x=1mm,y=1mm]\filldraw[fill=white,draw=white,thick](0,0) rectangle (35,30);\end{tikzpicture}}
	\put(120,98){\Large$\begin{array}{l}\lhcb \\ 9\invfb \end{array}$}
	\put(75,1){\LARGE$m_{\PX}$} 
	\put(119,0){\LARGE$\left[\!\gevcc\right]$}
	\put(-3,48){\LARGE\begin{sideways}$F_{\PX}\ \left[\mathrm{90\%CL~UL}\right]$\end{sideways}}
	\put(90,75){$\begin{array}{cl} 
	{\color{blue}{\rule[2pt]{8mm}{1.0pt}}}
      & \mathrm{observed}
      \\
     {\color{red}{\hdashrule[2pt]{8mm}{0.5pt}{0.4mm 0.4mm}}}
      &  \mathrm{expected}
      \\
	\begin{tikzpicture}[x=1mm,y=1mm]\filldraw[fill=green,draw=green,thick]  (0,0) rectangle (8,3);\end{tikzpicture} 
      & \mathrm{expected}\pm\!1\upsigma
      \\
     \begin{tikzpicture}[x=1mm,y=1mm]\filldraw[fill=yellow,draw=yellow,thick]  (0,0) rectangle (8,3);\end{tikzpicture} 
      & \mathrm{expected}\pm\!2\upsigma
      \\
  	\begin{tikzpicture}[x=1mm,y=1mm]\filldraw[fill=brightturquoise,draw=brightturquoise,thick]  (0,0) rectangle (8,3);\end{tikzpicture} 
      & \mathrm{expected}\pm\!3\upsigma
 
	\end{array}$}
	\end{picture}
	\caption {\small
	Upper limit\,(90\%~CL) on the~ratio of branching fractions 
	$F_{\PX}$ 
	as a~function of the~mass  
	of the~hypothetical narrow \PX~state.
	The~median expected upper limit together with 
	the~expected CL bands corresponding
	to 1, 2 and 3 standard deviations are also shown.
	}
	\label{fig:upper_limit_hypo}
\end{figure}

\begin{figure}[t]
	\setlength{\unitlength}{1mm}
	\centering
	\begin{picture}(170,120)
	\put(0,1){\includegraphics*[width=150mm]{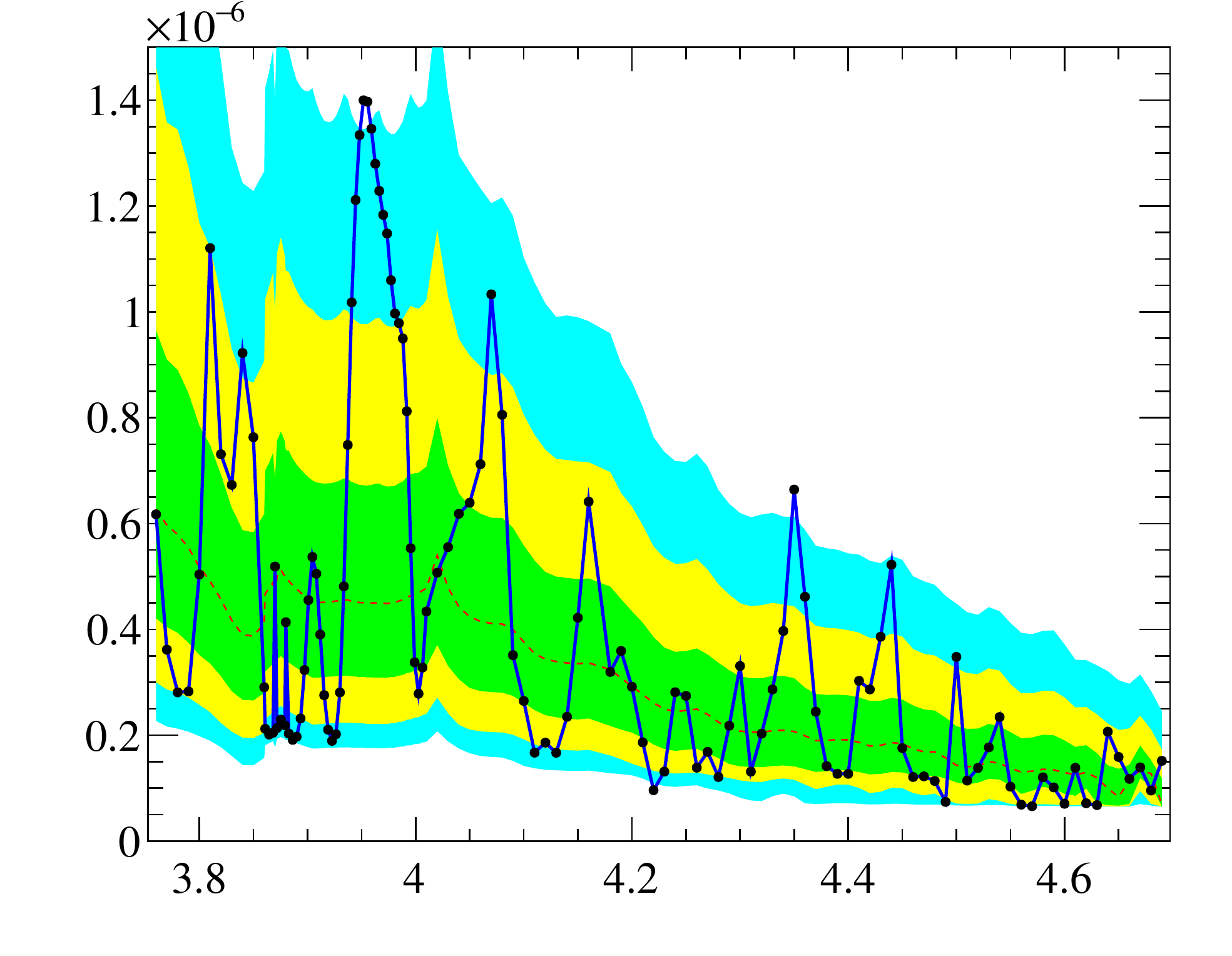}}
        \definecolor{aqua}{rgb}{0.0, 1.0, 1.0}
        \definecolor{brightturquoise}{rgb}{0.03, 0.91, 0.87}
	\put(120,98){\Large$\begin{array}{l}\lhcb \\ 9\invfb \end{array}$}

	\put(75,1){\LARGE$m_{\PX}$} 
	\put(119,1){\LARGE$\left[\!\gevcc\right]$}
	\put( 1,63){\LARGE\begin{sideways}$B_{\PX}\ \left[\mathrm{90\%CL~UL}\right]$\end{sideways}}
	\put(90,75){$\begin{array}{cl} 
	{\color{blue}{\rule[2pt]{8mm}{1.0pt}}}
      & \mathrm{observed}
      \\
     {\color{red}{\hdashrule[2pt]{8mm}{0.5pt}{0.4mm 0.4mm}}}
      &  \mathrm{expected}
      \\
	\begin{tikzpicture}[x=1mm,y=1mm]\filldraw[fill=green,draw=green,thick]  (0,0) rectangle (8,3);\end{tikzpicture} 
      & \mathrm{expected}\pm\!1\upsigma
      \\
     \begin{tikzpicture}[x=1mm,y=1mm]\filldraw[fill=yellow,draw=yellow,thick]  (0,0) rectangle (8,3);\end{tikzpicture} 
      & \mathrm{expected}\pm\!2\upsigma
      \\
  	\begin{tikzpicture}[x=1mm,y=1mm]\filldraw[fill=brightturquoise,draw=brightturquoise,thick]  (0,0) rectangle (8,3);\end{tikzpicture} 
      & \mathrm{expected}\pm\!3\upsigma
	\end{array}$}
	\end{picture}
	\caption {\small
	Upper limit\,(90\%~CL) on 
	the~product of branching fractions 
	$B_{\PX}$ 
	as a~function of the~mass  
	of the~hypothetical narrow \PX~state.
	The median expected upper limit together with 
	the~expected CL bands corresponding
	to 1, 2 and 3 standard deviations are also shown.
	}
	\label{fig:upper_limit_hypo_2}
\end{figure}

The~upper limits at $90\%$~CL on
the~ratio of branching 
fractions $F_{\PX}(m_{\PX})$
for~\mbox{$\decay{\Bu}
{\left( \decay{\PX}{\jpsi\Peta}\right) \Kp}$}~decays 
via a~narrow intermediate $\PX$~state
are set for masses 
of the~hypothetical  \PX~particle 
between 3.7 and 4.7\gevcc. 
The~upper limits, 
shown in Fig.~\ref{fig:upper_limit_hypo}, 
are set with the~$CL_s$ method~\cite{CLs} in which 
the~$p$\nobreakdash-values
are calculated based on 
the~asymptotic properties
of the~profile likelihood ratio~\cite{Asymptoticformulae}.
The~corresponding upper limits 
on the~product of branching fractions, $B_{\PX}$, 
are calculated in 
a~similar way 
and shown  in Fig.~\ref{fig:upper_limit_hypo_2}.
The~local statistical significance 
for the~mass values 
with the~weakest upper limits,
\eg $m_{\PX}=3.952$, $4.352$ or \mbox{$4.442\gevcc$},
is estimated using Wilks' theorem~\cite{Wilks:1938dza}
and  
is found to be less than three~standard deviations.

Signals with a~statistical significance exceeding 
three standard deviations are seen only for 
the~\mbox{$\decay{\Bu}
{\left(\decay{\Ppsi_2(3823)}{\jpsi\Peta}\right)\Kp}$}
and \mbox{$\decay{\Bu}
{\left( \decay{\Ppsi(4040)}
{\jpsi\Peta}\right) \Kp}$}~decays.
The~fit to the~$\jpsi\Peta$~mass 
distribution 
in the~$\Ppsi(4040)$~region
suggests potential contributions 
from other resonances or 
sizeable interference effects,
in particular with a~possible 
$\Ppsi(4160)$~contribution. 
Accounting for systematic 
uncertainties the~significance is 
found to be 3.4~and 4.7~standard deviations 
for decays mediated by 
the~$\Ppsi_2(3823)$ 
and $\Ppsi(4040)$~states, respectively.
The~ratios of branching fractions are found to be
\begin{subequations}
\begin{eqnarray*}
F_{\Ppsi_2(3823)} & = &   
\left( 5.95\,^{\,+\,3.38}_{\,-\,2.55}\,\right) 
\times 10^{-2}\,,
\\
F_{\Ppsi(4040)} & = &   
\left( 40.6 \pm 11.2 \right) \times 10^{-2}\,.
\end{eqnarray*}
The~asymmetric uncertainty in $F_{\Ppsi_2(3823)}$ arises 
from varying the~$\Ppsi_2(3823)$ natural width between 0.2 and 6.6\mev.
The~corresponding products of branching fractions
are 
\begin{eqnarray*}   
B_{\Ppsi_2(3823)} & = & 
\left( 1.25 \,^{\,+\,0.71}_{\,-\,0.53} \pm 0.04 \right) 
\times 10^{-6} \,,
\\
B_{\Ppsi(4040)} & = &  
\left( 8.53 \pm 2.35 \pm 0.30\right) \times 10^{-6} \,,
\end{eqnarray*}
\end{subequations}
where the~second uncertainty is due to 
the~imprecise knowledge of 
the~\mbox{$\decay{\Bp}{\psitwos\Kp}$}
and \mbox{$\decay{\psitwos}{\jpsi\Peta}$}
branching fractions~\cite{PDG2021}.
 For~decays with other intermediate states
 no signals are seen and 
 the~corresponding upper limits 
 are listed in Table~\ref{tab:resonances2}. 

\begin{table}[t]
  \caption{
    \small 
    Upper limits at 90\%~CL for the~ratio of branching fractions $F_{\PX}$
    and product of branching fractions $B_{\PX}$
    for different conventional charmonia, charmonium\protect\nobreakdash-like and hypothetical states.  }
    \label{tab:resonances2}
\begin{center}
\begin{tabular*}{0.50\textwidth}{@{\hspace{3mm}}l@{\extracolsep{\fill}}cc@{\hspace{3mm}}}
     & \multicolumn{2}{c}{Upper limit at 90\%~CL }
    \\
     &  $F_{\PX}~\left[10^{-2}\right]$  
     &  $B_{\PX}~\left[10^{-7}\right]$  
    \\[2mm] 
    \hline
    \\[-2mm]
    $\Ppsi(3770)$     & $2.2$  &  $4.6$
    \\
    $\Ppsi_3(3842)$   &  2.9   &   6.1 
    \\
    $\Ppsi(4160)$     & $4.2$  &  $8.7$
    \\
    $\Ppsi(4415)$     & $4.6$  &  $9.6$
    \\[2mm]
    \hline
    \\[-2mm]
    $\PR(3760)$         & $2.0$    &  $ 4.1$  
    \\
    $\PR(3790)$         & $3.2$   &  $ 6.7$
    \\
    $\Z_{\cquark}(3900)^0$  & $2.1$  &  $4.3$
    \\
    $\Ppsi(4230)$     & $1.9$   &  $ 3.9$
    \\
    $\Ppsi(4360)$     & $6.0$   &  $ 12.4\phantom{0}$
    \\
    $\Ppsi(4390)$     & $11.6\phantom{0}$  &  $24.1\phantom{0}$
    \\[2mm]
    \hline
    \\[-2mm]
    $\PZ_{\cquark}(4430)^0$ & $6.1$  &  $12.7\phantom{0}$
    \\
    $\PX^{\prime}_{\PC}$  & $1.9$  &  $3.9$
    \end{tabular*}
    \end{center} 
    \end{table}

Using the~value of 
\mbox{$\BR(\decay{\Bu}{\Ppsi_2(3823)\Kp})
\times\BR(\decay{\Ppsi_2(3823)}{\jpsi\pip\pim})$}
from Ref.~\cite{LHCb-PAPER-2020-009},
the~ratio of branching fractions for 
the~\mbox{$\decay{\Ppsi_2(3823)}{\jpsi\Peta}$}
and \mbox{$\decay{\Ppsi_2(3823)}{\jpsi\pip\pim}$}
decays is calculated to be 
\begin{equation*}
    \dfrac  { \BR(
    \decay{\Ppsi_2(3823)}{\jpsi\Peta}) }
    { \BR( \decay{\Ppsi_2(3823)}{\jpsi\pip\pim}) } 
    = 
    4.4 \,^{\,+\,2.5}_{\,-\,1.9} \pm 0.9 
    \,, \, 
\end{equation*}
where the~last uncertainty accounts for 
the~precision  on external 
branching fractions~\cite{LHCb-PAPER-2020-009,PDG2021}.
Such a~large partial width of  
the~\mbox{$\decay{\Ppsi_2(3823)}{\jpsi\Peta}$}~decay
calls for a significant reevaluation 
of the~$\Ppsi_2(3823)$~branching 
fraction estimates of Ref.~\cite{Xu:2016kbn}.
This ratio is significantly larger than 
the~value of \mbox{$\left(9.72\pm0.14\right)
\times10^{-2}$} 
obtained for decays of the~$\psitwos$~state.
However, this might not be surprising as for
higher charmonium excitations 
the~charmonium\nobreakdash-to\nobreakdash-charmonium 
transitions with 
the~emission of an~\Peta~meson
are not suppressed, \eg for the~$\Ppsi(4040)$ state 
the~corresponding ratio 
of transitions involving $\Peta$
and $\pip\pim$ 
exceeds unity~\cite{PDG2021}.

Using the~previously measured value 
of the~\mbox{$\decay{\Ppsi(4040)}{\jpsi\Peta}$}
branching fraction 
from Refs.~\cite{BESIII:2012fdg,PDG2021},
the~\mbox{$\decay{\Bp}{\Ppsi(4040)\Kp}$}
branching fraction is calculated to be 
\begin{equation*}
 \BR( \decay{\Bp}{ \Ppsi(4040) \Kp }) 
 =\left ( 1.64 \pm 0.45 \pm  0.23 \right) \times 10^{-3}\,,
\end{equation*}
where
the~last uncertainty accounts for 
the~\mbox{$\decay{\Bp}{\psitwos\Kp}$},  
\mbox{$\decay{\psitwos}{\jpsi\Peta}$}
and 
\mbox{$\decay{\Ppsi(4040)}{\jpsi\Peta}$}~branching 
fraction uncertainties.
This disagrees with the~upper 
limit
of $1.3\times 10^{-4}$
at 90\%~CL
from Ref.~\cite{LHCb-PAPER-2013-039}, 
which used 
the~\mbox{$\decay{\Ppsi(4040)}{\mumu}$}~decay mode and 
relied on the 
\mbox{$\decay{\Ppsi(4040)}{\mumu}$}
branching fraction 
from Ref.~\cite{BES:2007zwq}
to obtain  a~\Bp~decay limit. 
This~disagreement motivates a~more 
detailed study of this system, 
such as a~full amplitude analysis.
In Ref.~\cite{LHCb-PAPER-2020-025} it was demonstrated 
that the~\mbox{$\decay{\Bp}
{\left( \decay{\Ppsi(4040)}{\Dp\Dm}\right)\Kp}$}
and \mbox{$\decay{\Bp}
{\left( \decay{\Ppsi(4160)}{\Dp\Dm}\right)\Kp}$}~decays
have comparable rates. 
In~this~paper 
a~significant suppression of 
the~\mbox{$\decay{\Ppsi(4160)}
{\jpsi\Peta}$}~transitions
relative to~such transitions 
for the~$\Ppsi(4040)$~state
is found.
The~opposite pattern was found in
the~analysis 
of~\mbox{$\decay{\Bp}
{\Kp\mumu}$}~decays~\cite{LHCb-PAPER-2013-039}, 
where a~large contribution from decays via 
the~intermediate 
$\Ppsi(4160)$~state is observed, 
while no 
decays with 
the~intermediate $\Ppsi(4040)$~state are seen. 

No signals are found for 
the~\mbox{$\decay{\Bp}{\jpsi\Peta\Kp}$}~decay via 
conventional charmonium states
$\Ppsi(3770)$,
$\Ppsi_3(3842)$, 
$\Ppsi(4160)$,
$\Ppsi(4415)$;
charmonium\nobreakdash-like states  
$\PR(3760)$,  
$\PR(3790)$,
$\Z_{\cquark}(3900)^0$,
$\Ppsi(4230)$,
$\Ppsi(4360)$ 
and $\Ppsi(4390)$;
the~hypothetical neutral partner
of the~charged 
$\PZ_{\cquark}(4430)^+$~state; 
and for the~hypothetical 
C\nobreakdash-odd partner 
of the~$\chicone(3872)$~state,
$\PX^{\prime}_{\PC}$.
In~particular, for the~latter 
upper limits at 90\%~CL are found to be
\begin{eqnarray*}
F_{\PX^{\prime}_{\PC}}       
      & < & 1.9\times 10^{-2} \,,
\\
B_{\PX^{\prime}_{\PC}}  
 & < & 3.9\times 10^{-7} \,,
\end{eqnarray*}
significantly improving 
results previously 
obtained
by the~\babar and
\belle~collaborations~\cite{BaBar:2004iez,
Belle:2013vio}. 

%
In conclusion, a~search for charmonium and 
charmonium\nobreakdash-like exotic 
states contributing to the~$\jpsi\Peta$~mass 
spectrum from~\mbox{$\decay{\Bp}
{\jpsi\Peta\Kp}$~decays} is performed,
using a~data sample 
corresponding to an~integrated 
luminosity of 9\invfb collected
 with the~LHCb detector
at 7, 8 and 13 \tev
centre\nobreakdash-of\nobreakdash-mass energies 
in proton\nobreakdash-proton collisions.
The~\mbox{$\decay{\Bp}
{\left(\decay{\psitwos}{\jpsi\Peta}\right)
\Kp}$}~decay mode is used  
a~normalisation channel. 
While no narrow resonances are seen, 
evidence is found for 
the~\mbox{$\decay{\Bp}
{\left(\decay{\Ppsi_2(3823)}{\jpsi\Peta}\right)\Kp}$}
and \mbox{$\decay{\Bp}
{\left(\decay{\Ppsi(4040)}{\jpsi\Peta}\right)\Kp}$}~decays, 
and the~corresponding branching fractions and their 
ratios relative to the~normalisation decay mode
are measured.

\section*{Acknowledgements}
%
%
\noindent We express our gratitude to our colleagues 
in the~CERN
accelerator departments for 
the~excellent performance of the~LHC. 
We~thank the~technical and administrative staff 
at the~LHCb institutes.
We~acknowledge support from CERN and from the national agencies:
CAPES, CNPq, FAPERJ and FINEP\,(Brazil); 
MOST and NSFC\,(China); 
CNRS/IN2P3\,(France); 
BMBF, DFG and MPG\,(Germany); 
INFN\,(Italy); 
NWO\,(Netherlands); 
MNiSW and NCN\,(Poland); 
MEN/IFA\,(Romania); 
MSHE\,(Russia); 
MICINN\,(Spain); 
SNSF and SER\,(Switzerland); 
NASU\,(Ukraine); 
STFC\,(United Kingdom); 
DOE NP and NSF\,(USA).
We~acknowledge the~computing resources that are provided by CERN, 
IN2P3\,(France), 
KIT and DESY\,(Germany), 
INFN\,(Italy), 
SURF\,(Netherlands),
PIC\,(Spain), 
GridPP\,(United Kingdom), 
RRCKI and Yandex LLC\,(Russia), 
CSCS\,(Switzerland), 
IFIN\nobreakdash-HH\,(Romania), 
CBPF\,(Brazil),
PL\nobreakdash-GRID\,(Poland) 
and NERSC\,(USA).
We~are indebted to the~communities behind 
the~multiple open\nobreakdash-source
software packages on which we depend.
Individual groups or members have received support from
ARC and ARDC\,(Australia);
AvH Foundation\,(Germany);
EPLANET, Marie Sk\l{}odowska\nobreakdash-Curie Actions and ERC\,(European Union);
A*MIDEX, ANR, Labex P2IO and OCEVU, and 
R\'{e}gion Auvergne\nobreakdash-Rh\^{o}ne\nobreakdash-Alpes\,(France);
Key Research Program of Frontier Sciences of CAS, CAS PIFI, CAS CCEPP, 
Fundamental Research Funds for the~Central Universities, 
and Sci. \& Tech. Program of Guangzhou\,(China);
RFBR, RSF and Yandex LLC\,(Russia);
GVA, XuntaGal and GENCAT\,(Spain);
the~Leverhulme Trust, the~Royal Society
 and UKRI\,(United Kingdom).

\clearpage 

\addcontentsline{toc}{section}{References}
\bibliographystyle{LHCb}
\bibliography{main,standard,LHCb-PAPER,LHCb-CONF,LHCb-DP,LHCb-TDR}

\newpage
\centerline
{\large\bf LHCb collaboration}
\begin
{flushleft}
\small
R.~Aaij$^{32}$,
A.S.W.~Abdelmotteleb$^{56}$,
C.~Abell{\'a}n~Beteta$^{50}$,
F.~Abudin{\'e}n$^{56}$,
T.~Ackernley$^{60}$,
B.~Adeva$^{46}$,
M.~Adinolfi$^{54}$,
H.~Afsharnia$^{9}$,
C.~Agapopoulou$^{13}$,
C.A.~Aidala$^{87}$
% ...

{\footnotesize \it

% ...
$^{9}$Universit{\'e} Clermont Auvergne, CNRS/IN2P3, LPC, Clermont-Ferrand, France\\
% ...
$^{13}$LPNHE, Sorbonne Universit{\'e}, Paris Diderot Sorbonne Paris Cit{\'e}, CNRS/IN2P3, Paris, France\\
% ...
$^{32}$Nikhef National Institute for Subatomic Physics, Amsterdam, Netherlands\\
% ...
$^{46}$Instituto Galego de F{\'\i}sica de Altas Enerx{\'\i}as (IGFAE), Universidade de Santiago de Compostela, Santiago de Compostela, Spain\\
% ...
$^{50}$Physik-Institut, Universit{\"a}t Z{\"u}rich, Z{\"u}rich, Switzerland\\
% ...
$^{54}$H.H. Wills Physics Laboratory, University of Bristol, Bristol, United Kingdom\\
% ...
$^{56}$Department of Physics, University of Warwick, Coventry, United Kingdom\\
% ...
$^{60}$Oliver Lodge Laboratory, University of Liverpool, Liverpool, United Kingdom\\
% ...
$^{68}$Syracuse University, Syracuse, NY, United States\\
% ...
$^{87}$University of Michigan, Ann Arbor, United States, associated to $^{68}$\\
\bigskip
% ...
}
\end{flushleft}

\end{document}